\documentclass[a4paper,fleqn,usenatbib]{mnras}

\usepackage[T1]{fontenc}
\usepackage{ae,aecompl}

\usepackage{graphicx}	% Including figure files
\usepackage{amsmath}	% Advanced maths commands
\usepackage{amssymb}	% Extra maths symbols
\usepackage{url}
\usepackage{gensymb}
\usepackage{subcaption}
\usepackage{float}
\usepackage{placeins}
\usepackage{soul}
\usepackage{ulem}

\usepackage[dvipsnames]{xcolor}

\usepackage{tikz}
\usepackage{graphicx}
\usepackage{tabularx} 

\usepackage [english]{babel}
\usepackage [autostyle, english = american]{csquotes}
\MakeOuterQuote{"}
\title[New VST ATLAS ISW measurements]{Cross-correlating Planck with VST ATLAS LRGs: a new test for the ISW effect in the Southern Hemisphere}

\author[B. Ansarinejad]{
Behzad Ansarinejad$^{1}$\thanks{E-mail: behzad.ansarinejad@durham.ac.uk}, Ruari Mackenzie$^{2}$, Tom Shanks$^{1}$, Nigel Metcalfe$^{1}$
\\
$^{1}$Department of Physics, Durham University, South Road, Durham, DH1 3LE, UK\\
$^{2}$Department of Physics, ETH Z{\"u}rich, Wolfgang-Pauli-Strasse 27, 8093 Z{\"u}rich, Switzerland
}

\date{Accepted: 2020 February 25; Revised: 2020 February 11; Received: 2019 September 25}

\pubyear{2020}

\begin{document}
\label{firstpage}
\pagerange{\pageref{firstpage}--\pageref{lastpage}}
\maketitle

\begin{abstract}
The Integrated Sachs-Wolfe (ISW) effect probes the late-time expansion history of the Universe, offering direct constraints on dark energy. Here we present our measurements of the ISW signal at redshifts of $\bar{z}=0.35$, $0.55$ and $0.68$, using the cross-correlation of the Planck CMB temperature map with $\sim0.5$ million Luminous Red Galaxies (LRGs) selected from the VST ATLAS survey. We then combine these with previous measurements based on WMAP and similar SDSS LRG samples, providing a total sample of $\sim2.1$ million LRGs covering $\sim12000$ deg$^2$ of sky. At $\bar{z}=0.35$ and $\bar{z}=0.55$ we detect the ISW signal at $1.2\sigma$ and $2.3\sigma$ (or $2.6\sigma$ combined), in agreement with the predictions of $\Lambda$CDM. We verify these results by repeating the measurements using the BOSS LOWZ and CMASS, spectroscopically confirmed LRG samples. We also detect the ISW effect in three magnitude limited ATLAS+SDSS galaxy samples extending to $z\approx0.4$ at $\sim2\sigma$ per sample. However, we do not detect the ISW signal at $\bar{z}=0.68$ when combining the ATLAS and SDSS results. Further tests using spectroscopically confirmed eBOSS LRGs at this redshift remain inconclusive due to the current low sky coverage of the survey. If the ISW signal is shown to be redshift dependent in a manner inconsistent with the predictions of $\Lambda$CDM, it could open the door to alternative theories such as modified gravity. It is therefore important to repeat the high redshift ISW measurement using the completed eBOSS sample, as well as deeper upcoming surveys such as DESI and LSST.
\end{abstract}


%%%%%%%%%%%%%%%%%%%%%%%%%%%%%%%%%%%%%%%%%%%%%%%%%%

%%%%%%%%%%%%%%%%% BODY OF PAPER %%%%%%%%%%%%%%%%%%

\section{Introduction}
\label{sec:Introduction}

Based on the latest observational evidence, the Universe is believed to be spatially flat (\citealt{Planck2016}) and undergoing a late-time accelerating state of expansion (\citealt{Riess1998};
\citealt{Alam2017}). In the current standard model of cosmology
$\Lambda$CDM, dark energy, parameterised as a cosmological constant
($\Lambda$), is believed to be the driving force behind this late-time accelerating expansion. Various alternatives to the cosmological constant have been proposed including modified gravity
(\citealt{Clifton2012}), scale-invariant (\citealt{Maeder2017}) or
spatially inhomogeneous cosmological models (see e.g.
\citealt{Dunsby2010}; \citealt{Racz2017}). As a consequence of the accelerated expansion of the Universe, cosmic microwave background (CMB) photons passing through gravitational potential wells, caused by large scale structure such as galaxy clusters, are left with a net gain of energy as the potential wells become shallower as the photons cross them. The opposite effect takes place as the photons pass through gravitational potential peaks (i.e. voids) with the photons undergoing a net loss of energy. The combination of these phenomena leads to secondary anisotropies on the CMB temperature map known as the (late-time) Integrated Sachs-Wolfe (ISW; \citealt{SW1967}) effect. The signature of the ISW effect can be observed as a non-zero signal in the cross-correlation between the distribution of foreground tracers of mass (such as galaxies) and the temperature of CMB, providing a direct probe of the late-time expansion of the Universe. 

Early attempts at measuring the ISW signal using the cross-correlation method, include an analysis of the COBE CMB map by \cite{Boughn2002} followed by detections of the signal using the WMAP CMB data, albeit often at relatively low to moderate levels of significance (\citealt{Scranton2003};  \citealt{Nolta2004}; \citealt{Boughn2004}; \citealt{Corasaniti2005}; \citealt{Padmanabhan2005}; \citealt{Giannantonio2006}; \citealt{Cabre2006}; \citealt{Rassat2007}; \citealt{Raccanelli2008}; \citealt{Granett2009}; \citealt{Bielby2010}; \citealt{Sawangwit2010}; \citealt{Kovacs2013}). Other studies have
however claimed detections in the range of $3-5\sigma$ (\citealt{Fosalba2003}; \citealt{Fosalba2004}; \citealt{Vielva2006}; \citealt{McEwen2006}; \citealt{Giannantonio2008}; \citealt{Ho2008};
\citealt{Granett2008}; \citealt{Giannantonio2012}; \citealt{Goto2012}; \citealt{PlanckISW2016}).

Another common approach in measuring the ISW signal is by stacking of voids and super-clusters. Similar to the cross-correlation method, studies using this approach have obtained detection significances ranging from low to moderate (\citealt{Granett2015};
\citealt{Kovacs2017}), to $3\sigma$ or higher (\citealt{Papai2011};
\citealt{Nadathur2016}; \citealt{Cai2017}; \citealt{PlanckISW2016}). Interestingly, a number of these studies have reported a signal with a higher amplitude than expected based on $\Lambda$CDM predictions.  

Here, we follow the work of \cite{Sawangwit2010} where the ISW analysis was performed on photometrically selected Luminous Red Galaxies (LRGs) from SDSS in the Northern Hemisphere. Three redshift-limited LRG samples were created allowing the measurement the ISW signal at redshifts of $\bar{z}=0.35$, $0.55$ and $0.68$. Although an ISW signal consistent with $\Lambda$CDM was detected at $\bar{z}=0.35$ and $0.55$, no such signal was detected at $\bar{z}=0.68$ albeit, as in the other two cases, the errors were significant. Given the implications of any ISW deviations from $\Lambda$CDM predictions, the lack of detection of the ISW signal at $\bar{z}=0.68$ in SDSS, is a particularly important topic for investigation using independent samples of LRGs. \cite{Sawangwit2010} also detected the ISW effect in three magnitude limited galaxy samples ($18<r<19$, $19<r<20$ and $20<r<21$), peaking in redshift at $z\approx0.20$, $0.27$ and $0.36$, providing some confirmation of the ISW measurements in the two lower redshift LRG samples but not in the third, highest redshift, sample.

In this work we measure the ISW signal in the cross-correlation of
similar samples of galaxies to those of  \cite{Sawangwit2010} but now selected from the VST ATLAS Survey (\citealt{Shanks2015}), with the Planck CMB temperature map (\citealt{Planck2016cmb}). The VST ATLAS survey has the advantage of covering large areas ($\sim4070$ deg$^2$) of the previously unexplored Southern sky, making it an ideal dataset for improving ISW constraints. This is because most of the available area in the North has already been covered by SDSS and, since the ISW signal weakens beyond $z\approx1$, there is limited option to increase the signal at larger distances. Indeed, with these Southern VST ATLAS data we may be approaching the upper limit to the significance of ISW detection due to cosmic variance in our limited `local' volume (see \citealt{Francis2010}).
 
 VST ATLAS is thus located  wholly in the Southern Hemisphere and
is split into two areas by the Galactic Plane. In the Northern Galactic Cap (NGC), the survey covers an area of $\sim1450 \deg^2$, while the Southern Galactic Cap (SGC) covers an area of $\sim2620\deg^2$. In these regions, the survey provides imaging data in $ugriz$ bands to similar depths as SDSS in the North, but with superior seeing. We shall use these data to select three LRG samples and three magnitude limited samples, closely analogous to those created by \cite{Sawangwit2010} using SDSS.

In order to test  our LRG selections, we shall first compare the angular auto-correlation functions of our VST ATLAS LRG samples to those of \cite{Sawangwit2011}. After cross-checking our photometric selections, we shall perform the ISW measurements and combine our results with those of \cite{Sawangwit2010}, to obtain better constraints on the ISW effect at each redshift. As a further verification of the SDSS ISW measurements at $\bar{z}=0.35$, $\bar{z}=0.55$ and $\bar{z}=0.68$ we repeat the measurements using the LOWZ and CMASS LRG samples from Data Release 12 (DR12; \citealt{BOSSDR12}) of the SDSS BOSS survey and the eBOSS DR14 LRG sample \citep{Prakash2016} respectively. Unlike photometrically selected samples, these spectroscopically confirmed samples do not suffer from contamination due to stars, or from galaxies outside the redshift range, making them ideal datasets for further testing the SDSS photometric ISW measurements, in particular. We note however, that while spectroscopic samples are not affected by stellar contamination or systematics related to photometric redshifts, they are not immune to targeting systematics which could introduce artificial correlations between the inferred density field and factors including stellar density, fiber collisions and observing conditions. In this work, when using spectroscopic samples, we account for these potential systematics by applying the BOSS/eBOSS weights (where available), as described in section~\ref{sec:XCF Method}.

To test the robustness of our ISW detections, we perform rotation tests similar to those previously implemented by \cite{Sawangwit2010} and \cite{Giannantonio2012}, where the ISW cross-correlation measurement is performed on incremental rotations of the LRG overdensity maps with respect to the CMB map to test for systematics. In their analysis, \cite{Sawangwit2010} found that in approximately 1 to 2 out of 8 cases, the rotated maps produce a more significant ISW detection than the un-rotated map. Using a similar approach, \cite{Giannantonio2012} claimed  that the results of their rotation tests, were consistent with the statistical variance of their associated datasets. Here we shall apply the rotation test to the ISW measurements obtained from the BOSS LOWZ and CMASS spectroscopic LRG samples, to check their robustness and compare our findings with those of \cite{Giannantonio2012}.

Hence, our aims are first to use ATLAS to test the reproducibility of the ISW measurements in the three LRG and the three magnitude limited galaxy samples as selected by \cite{Sawangwit2010} in SDSS. Of particular interest, is whether the VST ATLAS data independently reproduce the null detection of the ISW effect in the highest redshift LRG sample at $\bar{z}=0.68$. Our second aim is to check the robustness of the previous SDSS LRG results using new spectroscopically confirmed SDSS LRG samples, particularly in the two lower redshift ranges. The final aim is to apply the rotation test to the BOSS LRG samples to assess the robustness of such ISW measurements. 

The layout of this paper is as follows: we present a description of the selected datasets in Section~\ref{sec:Datasets}, followed by an outline of all relevant methodology in Section~\ref{sec:Methodology}. We present the results of our analysis and a discussion of our findings in Section~\ref{sec:Results} and conclude this work in Section~\ref{sec:Conclusions}.

Throughout this work all magnitudes are given in the AB system, and for consistency, we assume the fiducial $\Lambda$CDM cosmology adopted by \cite{Sawangwit2010} with $\Omega_\Lambda=0.7$, $\Omega_{\textup{m}}=0.3$, $f_{\textup{baryon}}=0.167$, $\sigma_8=0.8$ and $h=0.7$.    

\section{Datasets}
\label{sec:Datasets}

\subsection{Planck 2016 CMB temperature map}
\label{sec:CMB_data}

\begin{figure*}
	\begin{subfigure}{\columnwidth}
		\includegraphics[width=\columnwidth]{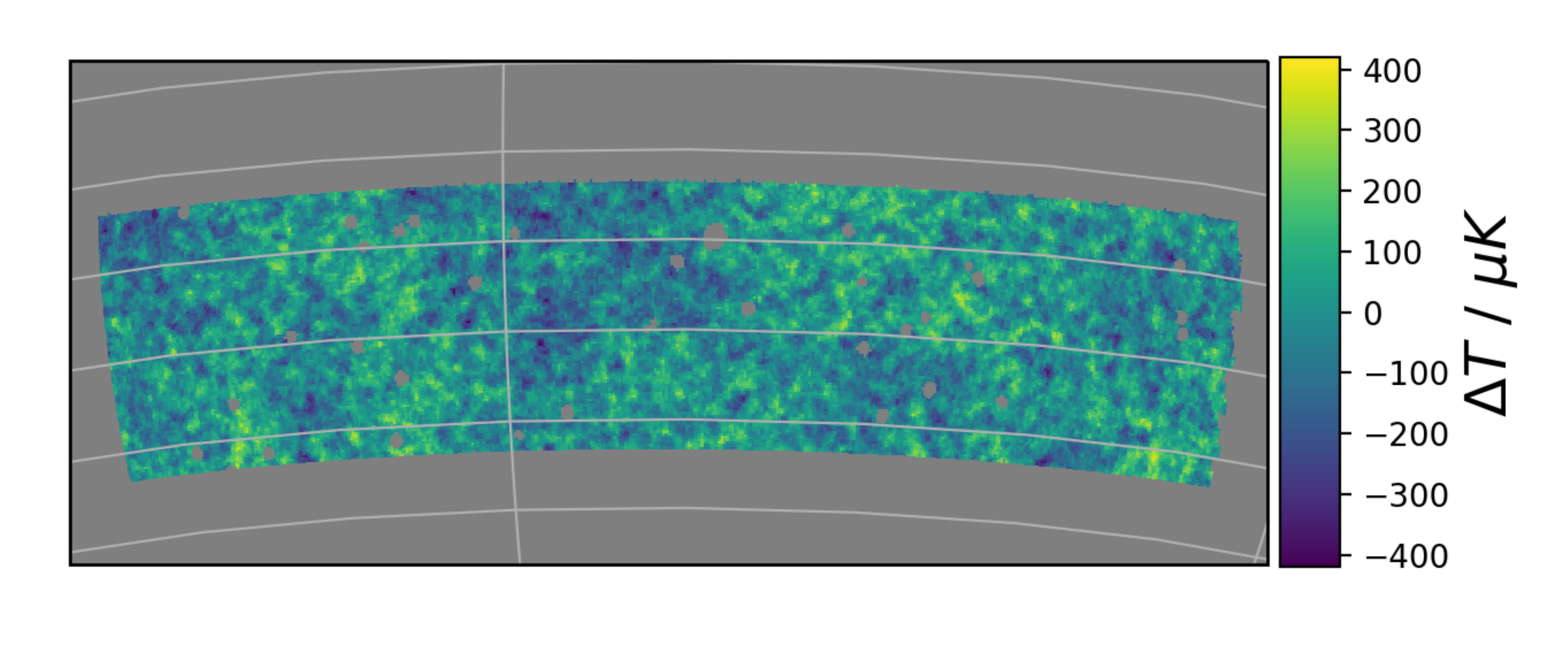} 
		\label{fig:cmb_NGC}
		\vspace{-2\baselineskip}
		\caption{CMB NGC}
	\end{subfigure}
		\begin{subfigure}{\columnwidth}
		\includegraphics[width=\columnwidth]{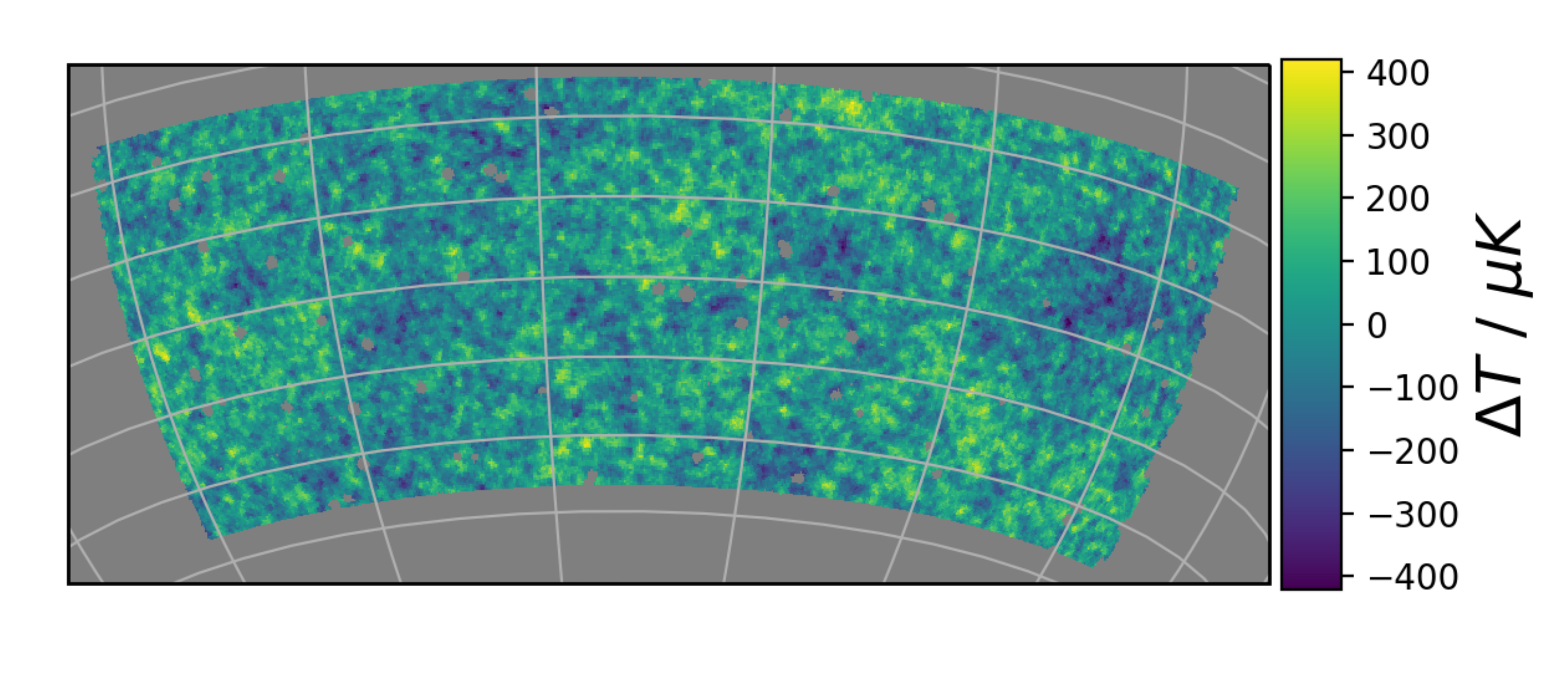} 
		\label{fig:cmb_SGC}
		\vspace{-2\baselineskip}
		\caption{CMB SGC}
	\end{subfigure}
		\begin{subfigure}{\columnwidth}
		\includegraphics[width=\columnwidth]{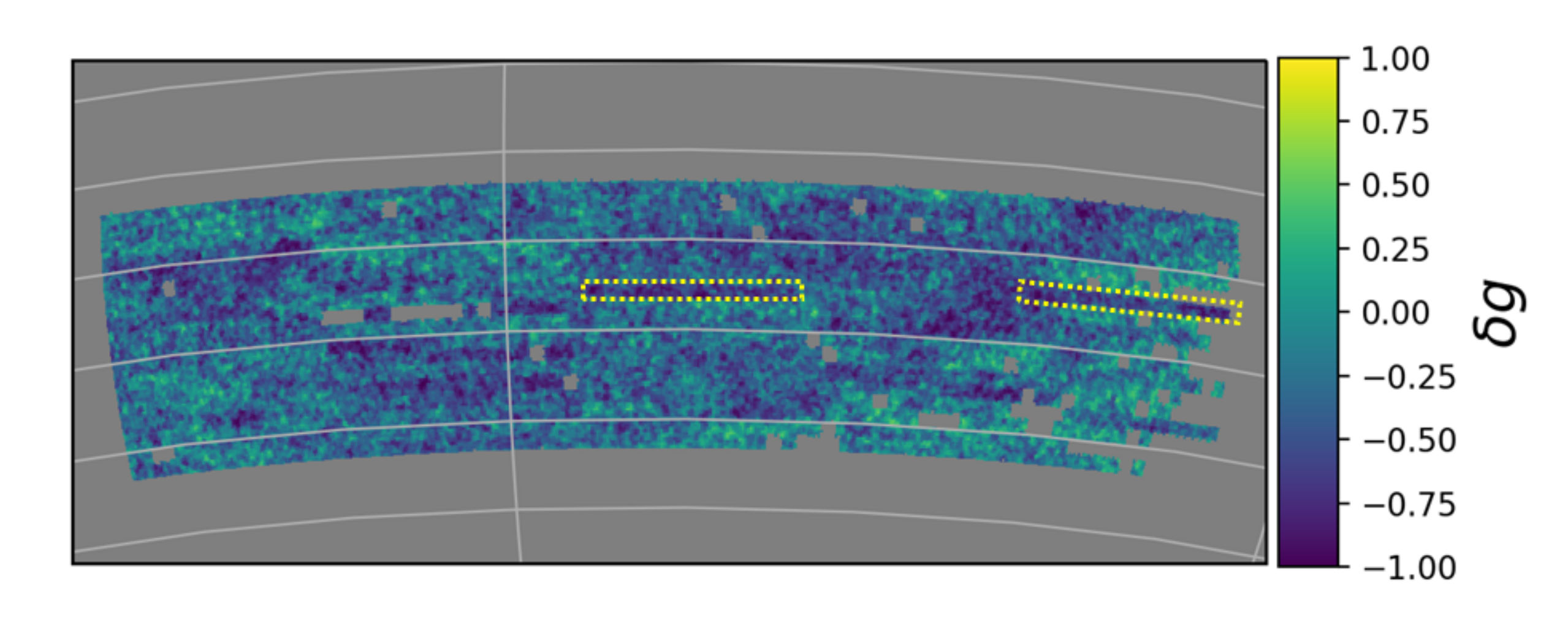} 
		\label{fig:AAO_NGC}
		\vspace{-2\baselineskip}
		\caption{$\bar{z}=0.68$ NGC}
	\end{subfigure}
		\begin{subfigure}{\columnwidth}
		\includegraphics[width=\columnwidth]{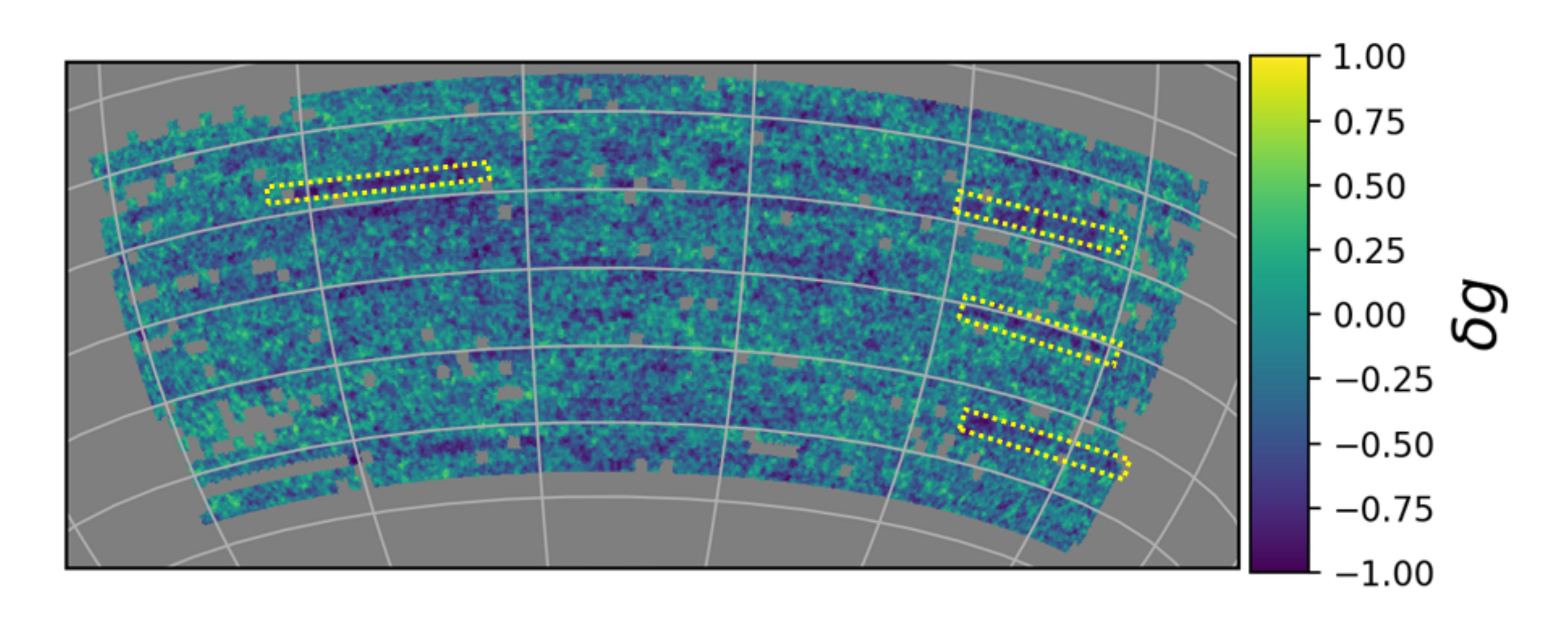} 
		\label{fig:AAO_SGC}
		\vspace{-2\baselineskip}
		\caption{$\bar{z}=0.68$ SGC}
	\end{subfigure}
		\begin{subfigure}{\columnwidth}
		\includegraphics[width=\columnwidth]{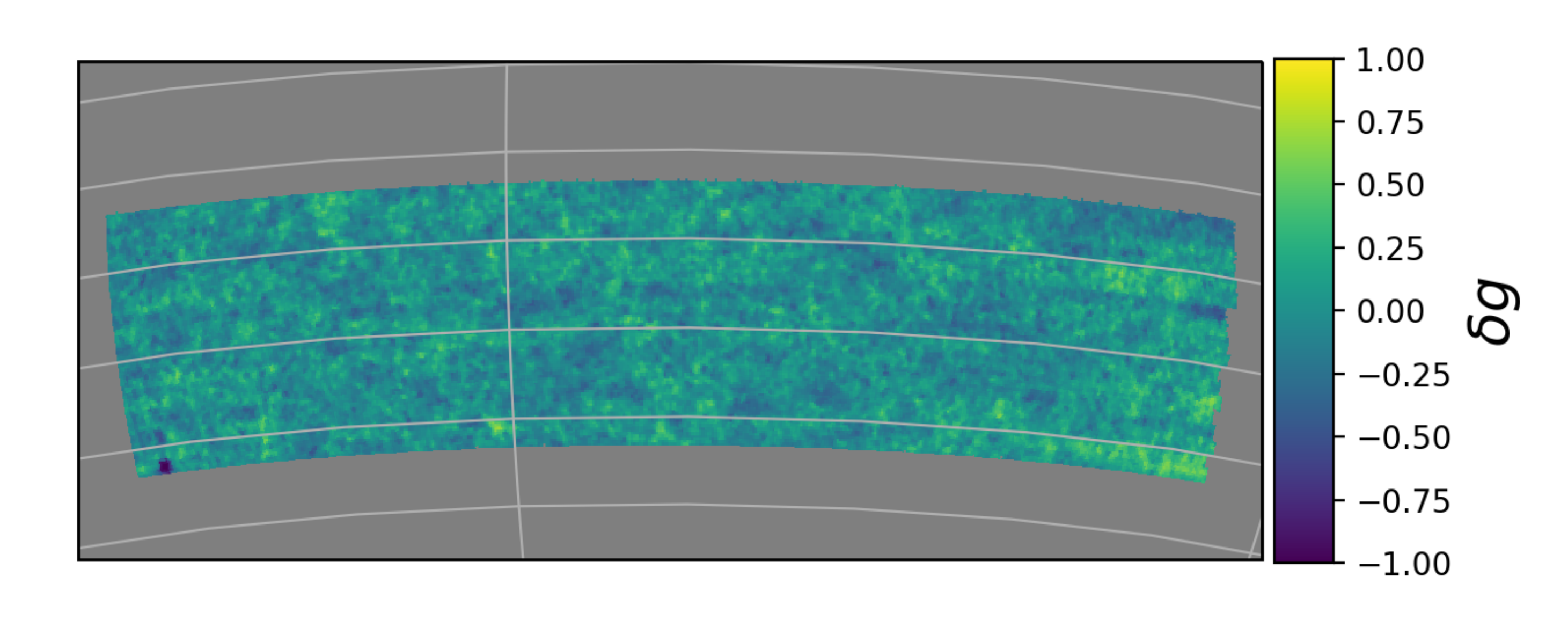} 
		\label{fig:20-21_NGC}
		\vspace{-2\baselineskip}
		\caption{20<r<21 NGC}
	\end{subfigure}
		\begin{subfigure}{\columnwidth}
		\includegraphics[width=\columnwidth]{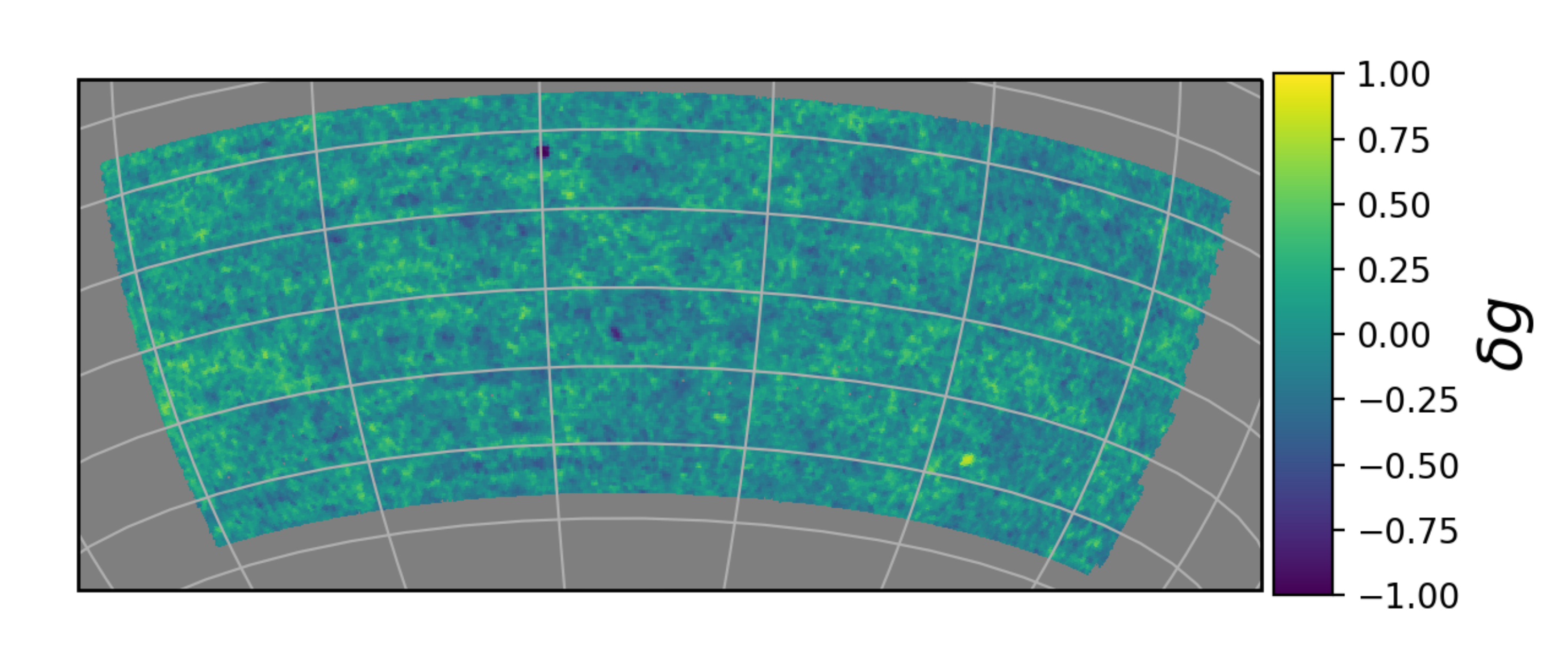} 
		\label{fig:20-21_SGC}
		\vspace{-2\baselineskip}
		\caption{20<r<21 SGC}
	\end{subfigure}
	\caption[]{The Planck CMB temperature anisotropy map covering the NGC (a) and SGC (b) of the VST ATLAS survey (described in Section~\ref{sec:CMB_data}). (c \& d) The VST ATLAS $\bar{z}=0.68$ LRG overdensity maps. Although we have included a number of under-dense concatenations in the final sample (marked by dotted boxes), our tests show that our auto and cross-correlation measurements are insensitive to masking these. (e \& f) The VST ATLAS galaxy overdensity maps for our $20<r<21$ magnitude limited sample (described in Section~\ref{sec:mag_lim}).}
	\label{fig:cmb_heatmap}
\end{figure*}

In our ISW analysis, we use the full Planck 2016 \texttt{COMMANDER} CMB temperature map (described in \citealt{Planck2016cmb}), downgraded to a \textsc{healpix}\footnote{http://healpix.sourceforge.net} (\citealt{Gorski2005}) resolution of $N_{side}=512$ (FWHM=20 arcmin). This is consistent with the \textsc{HEALpix} resolution used in the analysis of \cite{Sawangwit2010}. We apply the associated \texttt{COMMANDER} "confidence" mask to remove sections of the sky where the temperature and polarization CMB solution cannot be trusted. The masked Planck CMB maps corresponding to coverage area of VST ATLAS Northern and Southern Galactic Caps shown in Figure~\ref{fig:cmb_heatmap}. In this Figure, we also show the overdensity maps for our $\bar{z}=0.68$ LRG and $20<r<21$ magnitude limited galaxy sample, as described in Sections~\ref{sec:z0.68} and~\ref{sec:mag_lim} respectively. Here we show the overdensity maps of our faintest and highest redshift samples, as they are the most challenging to obtain due to their susceptibility to residual stellar contamination and artificial inhomogeneities caused by factors such as varying observing conditions.

\subsection{VST ATLAS luminous red galaxies}
\label{sec:LRG_data}

The VST ATLAS catalogues are available from the Cambridge Astronomy Survey Unit (CASU\footnote{http://casu.ast.cam.ac.uk/}) and include $\sim5\sigma$ detections with Kron, Petrosian, fixed aperture fluxes and morphological classification for objects in each band as well as various other parameters\footnote{\label{note3}http://casu.ast.cam.ac.uk/surveys-projects/vst/technical/catalogue-generation}. Details of the VST ATLAS calibration can be found in \cite{Shanks2015} and improved global photometric calibration based on the Gaia survey \citep{Gaia2018} is provided with the fourth Data release (DR4) of the survey\footnote{https://www.eso.org/sci/publications/announcements/ sciann17211.html}. 

Unlike SDSS, model magnitudes are not currently available for the VST ATLAS survey and we utilize aperture magnitudes in defining our LRG photometric selections. We denote aperture magnitudes corresponding to the ATLAS \texttt{Aperture flux 3} and \texttt{Aperture flux 5} using subscripts `A3' and `A5'. These apertures have radii of $1$ and $2''$ respectively and we apply their associated aperture corrections labelled as \texttt{APCOR} in the CASU catalogue$^{\ref{note3}}$. For $g, r, i$ and $z$ bands, the mean values of \texttt{APCOR3} are 0.45, 0.42, 0.35 and 0.38 mags, while mean values of \texttt{APCOR5} are 0.12, 0.12, 0.11 and 0.12 mags. Although these aperture corrections are derived for stars, they also provide a first order seeing correction for faint galaxies, and overall, we find that aperture magnitudes appear to give the most consistent galaxy colours compared to SDSS model magnitudes. Where Kron magnitudes are used, we correct these to total magnitude for galaxies, based on the offset between a the ATLAS Kron and SDSS model magnitudes, adding corrections of $-0.28$ and $-0.35$mag to ATLAS $r_{Kron}$ and $i_{Kron}$ respectively. We correct all magnitudes for Galactic dust extinction $A_x=C_xE(B-V)$, with $x$ representing a filter ($griz$), taking the SDSS $C_x$ values presented in \cite{Schneider2007} (3.793, 2.751, 2.086, and 1.479 for $griz$ respectively) and using the Planck $E(B-V)$ map (\citealt{Planck2014}). 

Following the photometric selection criteria of \cite{Sawangwit2010}, which was used to extract LRGs from the SDSS Data Release 5 (DR5; \citealt{Adelman-McCarthy2007}) data, we use the VST ATLAS survey to define three LRG samples at low ($\bar{z}=0.35$), intermediate ($\bar{z}=0.55$) and high ($\bar{z}=0.68$) redshifts. \cite{Sawangwit2010} in turn adopted their selection criteria based on those of the SDSS LRG (\citealt{Eisenstein2001}), 2DF-SDSS LRG and QSO (2SLAQ; \citealt{Cannon2006}) and Anglo-Australian Telescope (AAT)-AAOmega (\citealt{Ross2008}) spectroscopic redshift surveys, corresponding to the low, intermediate and high redshift LRG samples respectively. In Appendix~\ref{sec:Appendix_A} we compare the ATLAS and SDSS $g-r$, $r-i$ and $i-z$ colours, finding a reasonably tight scatter with no major systematic offsets in all cases. This enables us to adopt the above mentioned SDSS-based photometric selection criteria in defining our redshift limited ATLAS LRG samples. Furthermore, we remove objects located close to bright stars by matching to the Tycho-2 bright star catalogue (\citealt{Hog2000}). This is done to mask the halos formed in these regions due to reflections from bright stars which could be misclassified as galaxies, when source extraction is performed on the images. Based on visual inspection of these halos we systematically mask circular regions around the stars with radii depending on the stars' $V_T$ magnitudes: $V_T\leq8 : 340$ arcsecs ; $8<V_T\leq9 : 80$ arcsecs ; $9<V_T\leq10 : 45$ arcsecs ; $10<V_T\leq11 : 30$ arcsecs ; $V_T>11 : 20$ arcsecs.

As observations for the VST ATLAS survey are taken in one band at a time and the telescope has a 1 deg$^2$ field of view (henceforth referred to as a `tile'), it is possible that different bands are observed on separate nights with varying atmospheric conditions. Although ATLAS has a relatively tight seeing distribution, in a small number of cases, variations in seeing could result in fewer objects being detected in one band (especially at fainter magnitudes). This is because unlike SDSS, forced photometry is currently unavailable for the VST ATLAS catalogue and in this work we 
only have access to colours for objects with $>5\sigma$ detection in each band. Consequently, when the selection is applied, regions of the sky covered by these tiles will appear under-dense. Conversely, a few tiles could have a much higher than average number density due to residual stellar contamination (particularly in the NGC where the edge of the survey approaches the Galactic plane). 

In order to reduce the impact of these factors on our clustering measurements, we impose a lower and an upper limit on the number of objects per tile, which masks any significantly under and over-dense tiles. This ensures the LRG samples used in our cross-correlation analysis do not contain artificial inhomogeneities due to photometric artefacts or residual stellar contamination. We select these lower and upper limits based on comparing the auto-correlation function of the LRG samples to the measurements of \cite{Sawangwit2011}, thus ensuring that such artefacts and contaminations do not impact our ability to recover the true clustering of the LRGs. 

\subsubsection{$\bar{z}=0.35$ low redshift LRG sample}
\label{sec:z0.35}

\begin{figure}
	\includegraphics[width=\columnwidth]{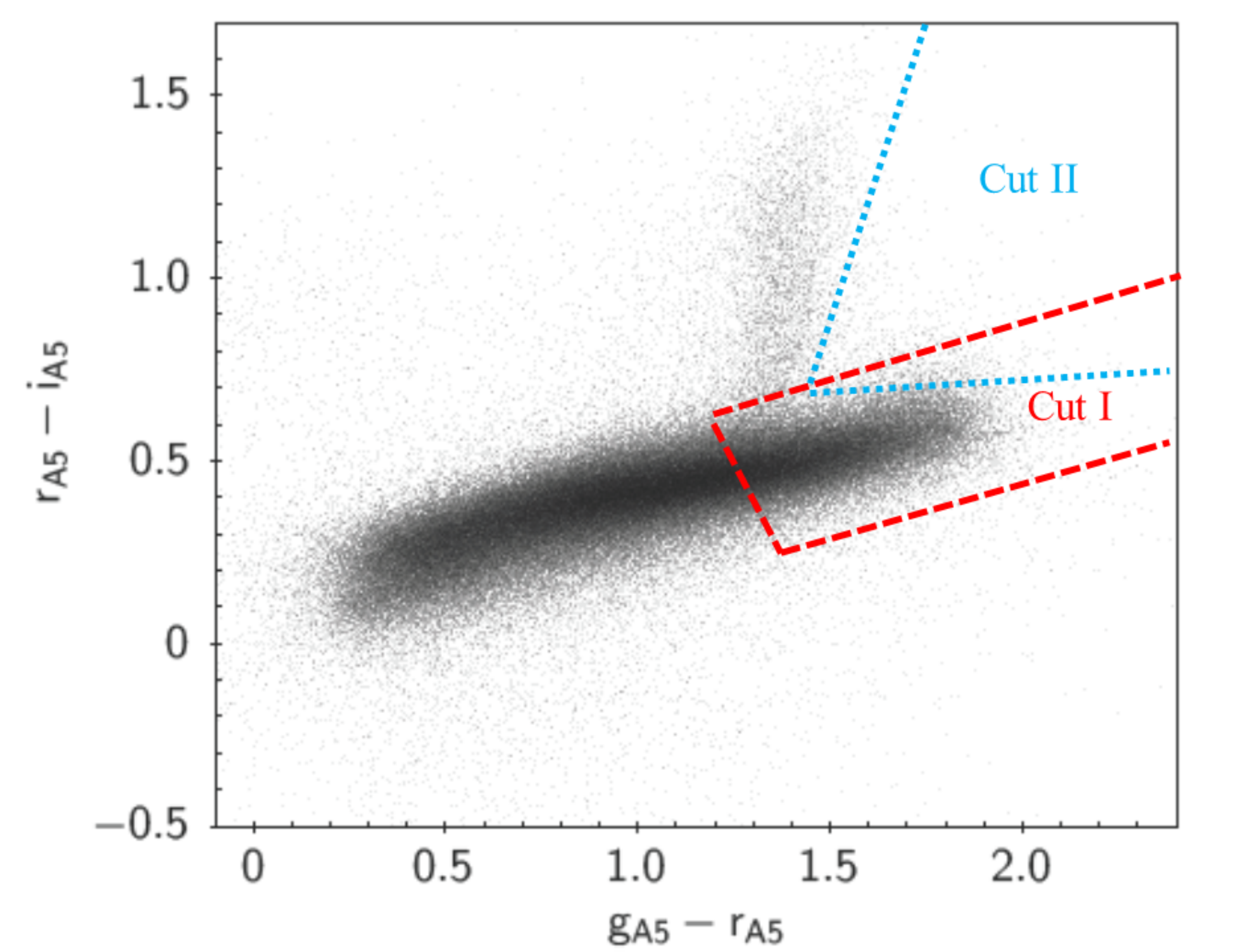}
	\caption{Our $\bar{z}=0.35$ LRG photometric selection, based on 'Cut I' and 'Cut II' of \protect\cite{Eisenstein2001}, used in photometric selection of SDSS LRGs. Here, as in subsequent plots, the colour gradient illustrates the density of the points, with darker shades representing a higher number of data points occupying a region of the colour space. The objects shown in this plot are classified as galaxies based on their VST ATLAS r-band morphological classification and lie within a magnitude limit of $17.5<r_{Kron}<19.5$.}
	\label{fig:sdss_cuts} 
\end{figure}

Objects in our low redshift LRG sample are selected based on satisfying the following conditions: 
\begin{gather}
17.5<r_{Kron}<19.2 , \\
r_{Kron}<13.1+c_{\parallel}/0.3 , \\
c_{\perp}<0.2 ,
\end{gather} corresponding to 'Cut I' of \cite{Eisenstein2001}, or 'Cut II' of the same study as defined by:
\begin{gather}
17.5<r_{Kron}<19.5 , \\
c_{\perp}>0.45-(g_{A5}-r_{A5})/6 , \\
g_{A5}-r_{A5}>1.3+0.25(r_{A5}-i_{A5}). 
\end{gather} The colour variables $c_{\parallel}$ are and $c_{\perp}$ are given by:
\begin{gather}
c_{\parallel}= 0.7(g_{A5}-r_{A5})+ 1.2(r_{A5}-i_{A5}-0.18), \label{eq:c_par} \\ 
c_{\perp}= (r_{A5}-i_{A5})-(g_{A5}-r_{A5})/4.0-0.18.
\end{gather}We note that our use of SDSS cuts in our LRG sample selection is justified given the similarity between ATLAS and SDSS bands (see Figure~\ref{fig:ATLAS_SDSS_colours} for comparison of ATLAS and SDSS colours). Figure~\ref{fig:sdss_cuts} shows the cuts used to selected our $\bar{z}=0.35$ LRG sample in the $r-i$ versus $g-r$ colour space.

\begin{figure}
	\includegraphics[width=\columnwidth]{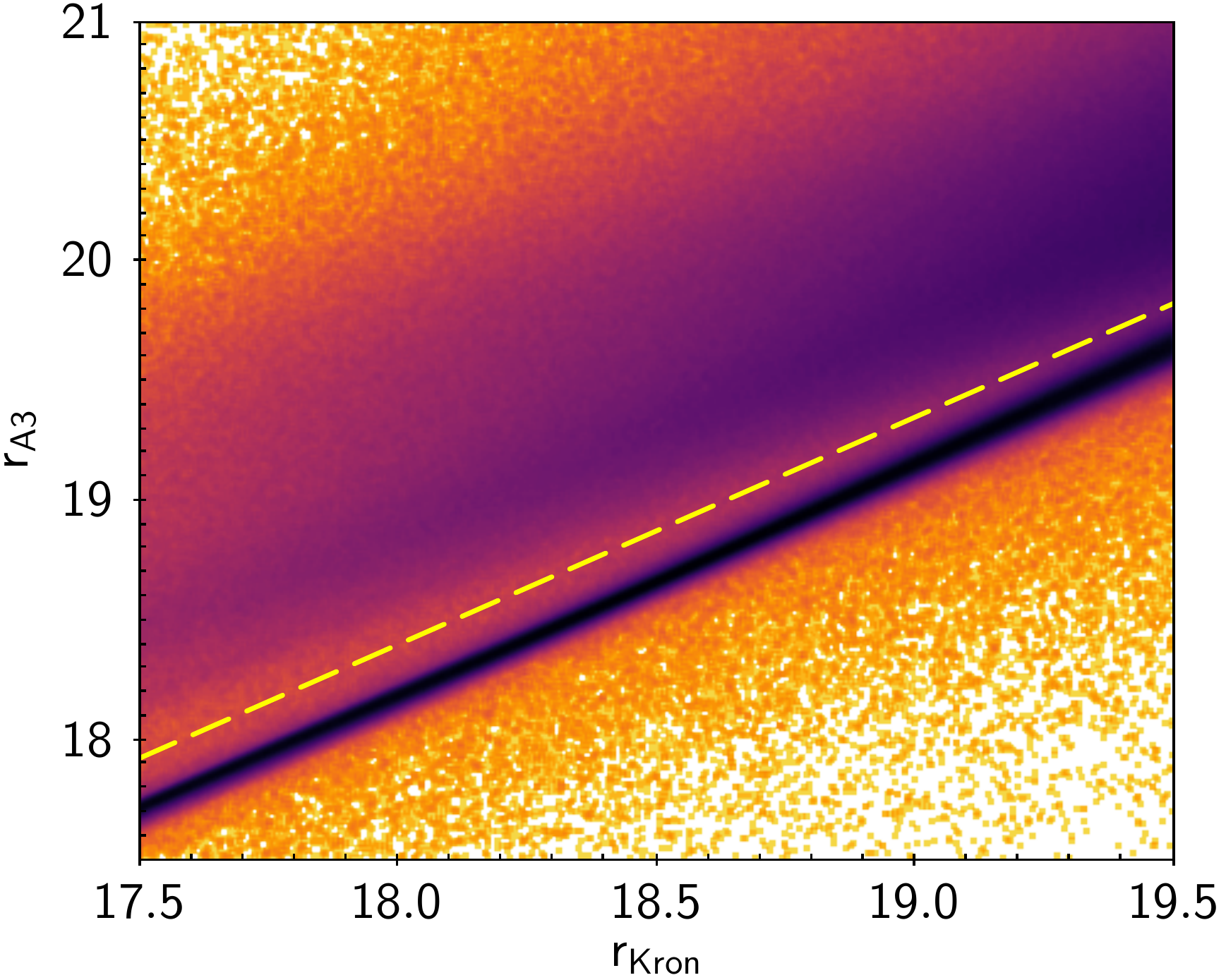}
	\caption{Further removal of residual stellar contamination from our $\bar{z}=0.35$ LRG sample. The dashed line is described by equation~\ref{eq:sdss_star_gal}, separating the stars (below the line) from galaxies (above the line).}
	\label{fig:sdss_star_gal} 
\end{figure}

To restrict our sample to galaxies we require the CASU r-band morphological classification \texttt{Classification\_r}=1, remove noisy regions (due to remaining ghost reflections from bright stars or large galaxies) by requiring \texttt{sky\_rms\_r}<0.2. We further remove residual stellar contamination via visual inspection of the $r_{A3}$ vs $r_{kron}$ diagram (see Figure~\ref{fig:sdss_star_gal}) by requiring: 
\begin{gather}
r_{A3}>0.909r_{kron}+2.
\label{eq:sdss_star_gal}
\end{gather}

In the case of our low redshift sample we mask tiles with fewer than 5 deg$^{-2}$ and more than 100 deg$^{-2}$ LRGs. This results in the removal of 10 tiles in the NGC and 8 in the SGC, leaving 31,531 ($\sim22$ deg$^{-2}$) and 63,245 ($\sim24$ deg$^{-2}$) LRGs in the NGC and SGC respectively.
 
\subsubsection{$\bar{z}=0.55$ intermediate redshift LRG sample}
\label{sec:z0.55}

\begin{figure}
	\includegraphics[width=\columnwidth]{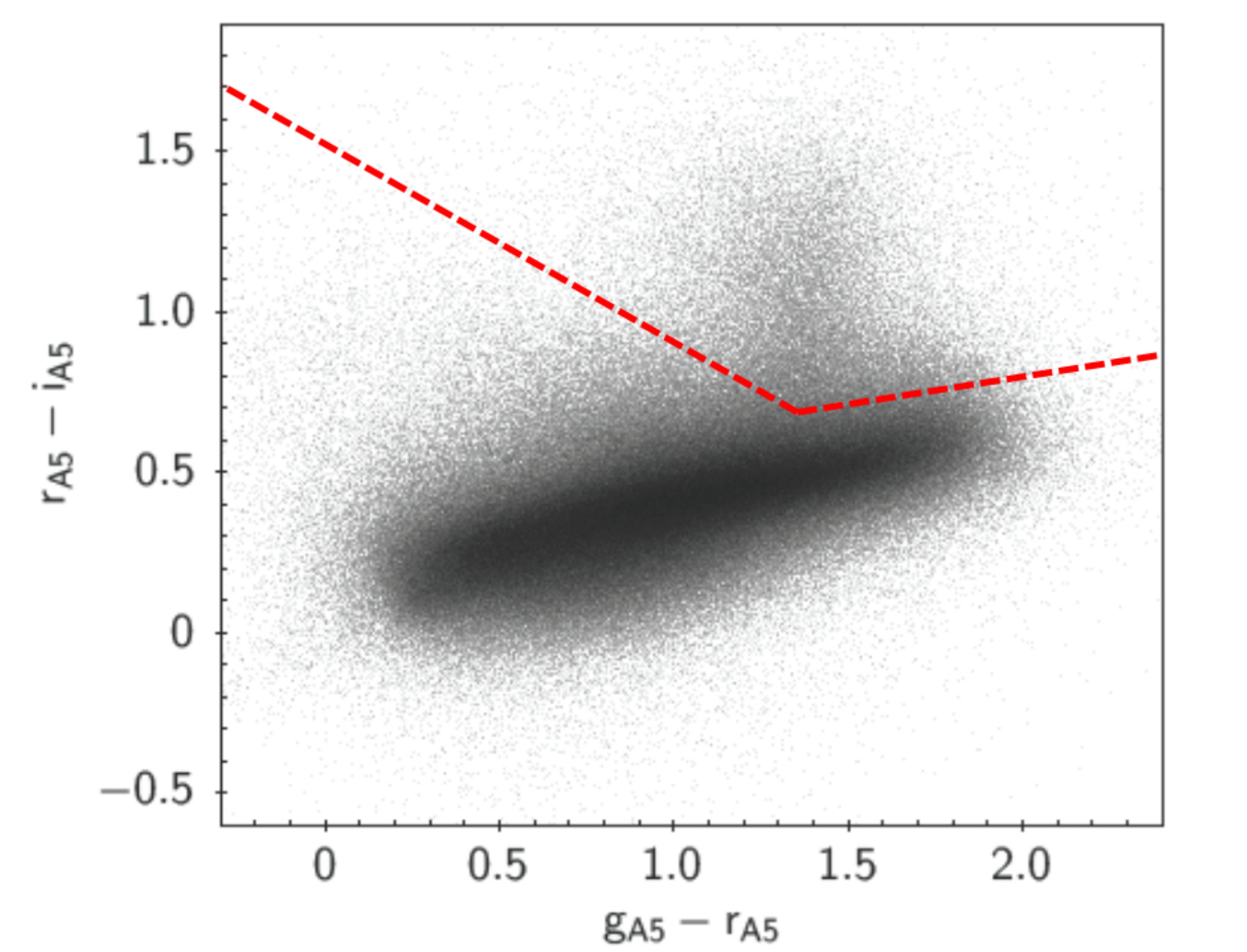}
	\caption{Our $\bar{z}=0.55$ LRG photometric selection, following the colour cuts of \protect\cite{Cannon2006} used in the selection of the 2SLAQ LRG sample. The objects shown here are classified as galaxies based on their VST ATLAS i-band morphological classification and lie within a magnitude limit of $17.5<i_{Kron}<19.8$.}
	\label{fig:2SLAQ_cuts} 
\end{figure}

We select our intermediate redshift LRG sample based on the following criteria which is an adaption of the photometric cuts of \cite{Cannon2006} used in the selection of the 2SLAQ LRG sample (see Figure~\ref{fig:2SLAQ_cuts}):
\begin{gather}
17.5\leq i_{Kron}<19.8 , \\
c_{\parallel}\geq 1.6 , \\
d_{\perp}>0.55 , \\
0.5 \leq(g_{A5}-r_{A5})\leq 3.0 , \\
(r_{A5}-i_{A5})<2 ,
\end{gather} where $c_{\parallel}$ is defined in equation~(\ref{eq:c_par}) and $d_{\perp}$ is given by: 
\begin{gather}
d_{\perp}=(r_{A5}-i_{A5})-(g_{A5}-r_{A5})/8.
\end{gather}

We restrict our selection to galaxies using the CASU i-band morphological classification (\texttt{Classification\_i}=1), limit \texttt{sky\_rms\_i}<0.2 and remove residual stellar contamination by imposing: 
\begin{gather}
i_{A3}>0.8i_{kron}+4.4,
\end{gather} in the NGC for $i_{kron}<19.6$ and in the SGC for $i_{kron}<19.1$, while imposing: 
\begin{gather}
i_{A3}>1.3i_{kron}-4.95 
\end{gather}for the NGC in the range $i_{kron}>19.6$ and
\begin{gather}
i_{A3}>1.8i_{kron}-14.7 
\end{gather} for the SGC in the range $i_{kron}>19.1$. Here we require different slopes for removing residual stars in the NGC compared to the SGC, as the edge of the survey lies closer to the Galactic plane in the NGC resulting in an increase in the level of contamination from residual stars.  

We mask tiles with fewer than 10 and more than 130 LRGs in the NGC and those with fewer than 10 and more than 150 LRGs in the SGC. This results in the removal of 21 tiles in the NGC and 8 tiles in the SGC, leaving 78,102 ($\sim55$ deg$^{-2}$) and 172,744 ($\sim66$ deg$^{-2}$) LRGs in the NGC and SGC respectively. 

\subsubsection{$\bar{z}=0.68$ high redshift LRG sample}
\label{sec:z0.68}

\begin{figure}
	\includegraphics[width=\columnwidth]{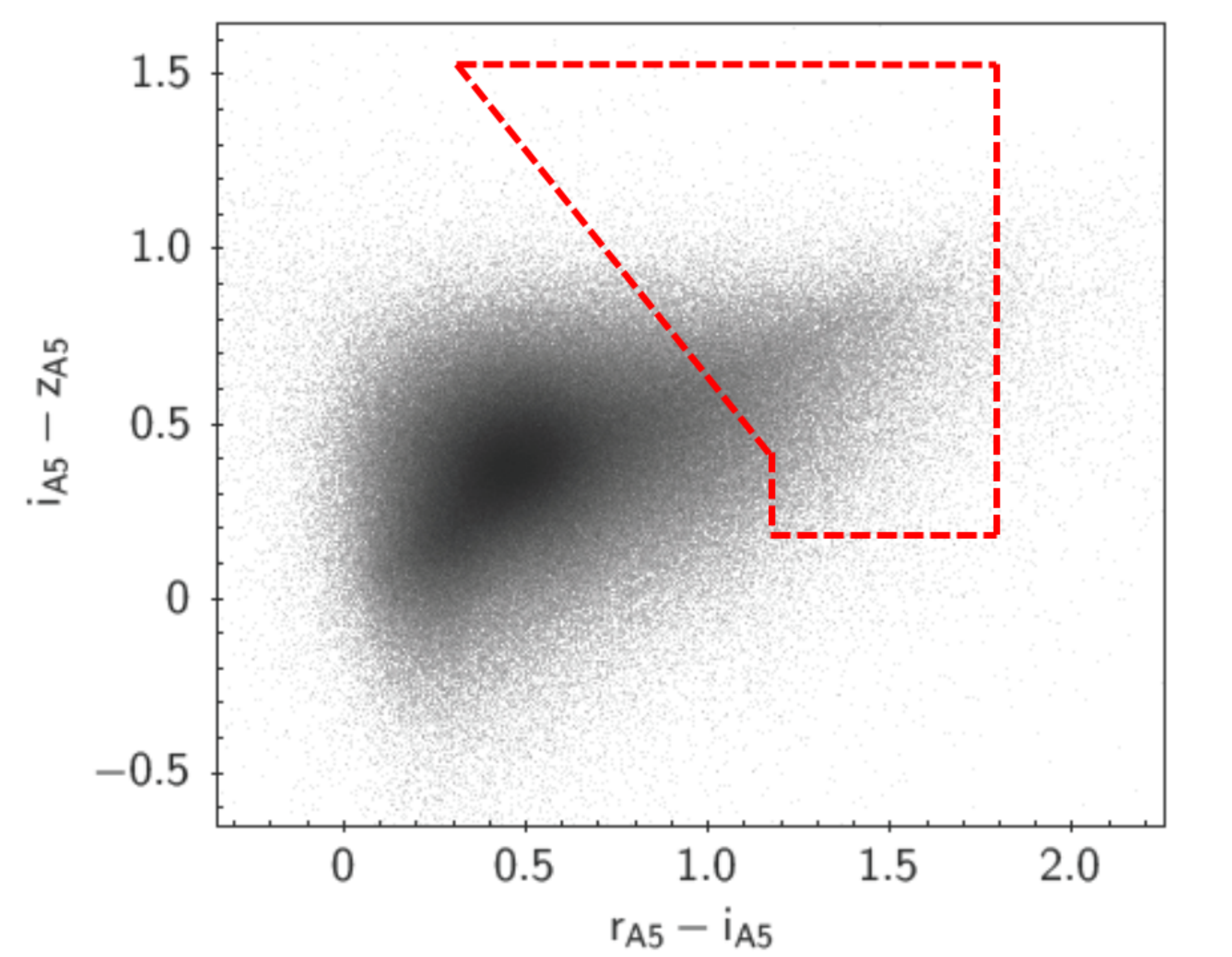}
	\caption{Our $\bar{z}=0.68$ LRG photometric selection, following the colour cuts of \protect\cite{Ross2008} used in the selection of the AAOmega LRG sample. The objects shown here are classified as galaxies based on their VST ATLAS i-band morphological classification and lie within a magnitude limit of $19.8<i_{Kron}<20.5$.}
	\label{fig:AAO_cuts} 
\end{figure}

The high redshift LRG sample is selected based on the following criteria (see Figure~\ref{fig:AAO_cuts}):
\begin{gather}
19.8<i_{Kron}<20.5 , \\
e_{\parallel}\geq 1.95 , \\
0.5\leq (r_{A5}-i_{A5})\leq 1.8 , \\
0.6\leq (i_{A5}-z_{A5})\leq 1.5 ,
\end{gather} or
\begin{gather}
0.2\leq (i_{A5}-z_{A5}) \leq 0.6, \\
x \leq (r_{A5}-i_{A5}) \leq 1.8, 
\end{gather} with $x$ being the smaller of $e_{\parallel}=(i_{A5}-z_{A5})+(9/7)(r_{A5}-i_{A5})$ or 1.2 at a given $(i_{A5}-z_{A5})$. The sample is restricted to galaxies using the CASU i-band morphological classification (\texttt{Classification\_i}=1) and stellar contamination is removed by imposing:
\begin{gather}
i_{A3}>1.2i_{kron}-3.45,
\end{gather} in the NGC for $i_{kron}<20.02$, otherwise: 
\begin{gather}
i_{A3}>1.4i_{kron}-7.42.
\end{gather} In the SGC the imposed cuts are: 
\begin{gather}
i_{A3}>1.2i_{kron}-3.55 , 
\end{gather} for $i_{kron}<20.23$, otherwise:
\begin{gather}
i_{A3}>1.4i_{kron}-7.55 .
\end{gather}

For this sample we mask tiles with fewer than 10 and more than 90 LRGs in both NGC and SGC. This excludes 101 tiles in the NGC and 192 in the SGC, leaving 62,379 ($\sim46$ deg$^{-2}$) and 138,977 ($\sim57$ deg$^{-2}$) LRGs in the NGC and SGC respectively. 

\begin{table*}
	\centering
	\caption{Details of the VST ATLAS, BOSS LOWZ and CMASS, and eBOSS LRG samples used in our cross-correlation analyses. For comparison we have included the same information for the SDSS LRG samples used in the analysis of \protect\cite{Sawangwit2010}. *Other magnitude limits used in the selection of the eBOSS sample can be found in \protect\cite{Prakash2016}.}
	\label{tab:LRG_samples}
	\begin{tabular}{rcccc} % four columns, alignment for each
		\hline
		Sample ($\bar{z}$) & Number of LRGs & Masked Area & Sky Density & Magnitude \\
		& & (deg$^2$) & ($\deg^{-2}$) & (AB) \\
		\hline
		ATLAS (0.35) & 94,776 & $\approx 4060$ & $\approx 23$ & $17.5< r<19.5$ \\
		ATLAS (0.55) & 250,846 & $\approx 4050$ & $\approx 62$ & $17.5< i<19.8$ \\
		ATLAS (0.68) & 201,356 & $\approx 3800$ & $\approx 53$ & $19.8< i<20.5$ \\
		\hline
		SDSS (0.35) & 106,699 & $\approx 8210$ & $\approx 13$ & $17.5< r<19.5$ \\
		SDSS (0.55) & 655,775  & $\approx 7715 $ & $\approx 85$ & $17.5< i<19.8$ \\
		SDSS (0.68) & 800,346 & $\approx 7622$ &  $\approx 105$ & $19.8< i<20.5$ \\
		\hline
		LOWZ (0.32) &  313,446 & $\approx 8337$ & $\approx 38$ & $16.0<r<19.6$ \\
		CMASS (0.57) & 849,637 & $\approx 9376$ & $\approx 91$ & $17.5< i<19.9$ \\
		eBOSS (0.70) & 141,000 & $\approx 1670$ & $\approx 84$ & $19.9< i<21.8$* \\
		\hline
	\end{tabular}
\end{table*}

\subsubsection{Magnitude limited galaxy samples}
\label{sec:mag_lim}

To select our three magnitude limited galaxy samples we require objects to be classified as galaxies using the CASU r-band morphological classification (\texttt{Classification\_r}=1). We then simply select objects satisfying $18<r_{kron}<19$, $19<r_{kron}<20$ and $19<r_{kron}<20$. In all cases we apply the additional cut of $r_{A5}>0.94r_{kron}+1.08$ to remove any residual stellar contamination from our samples. For the $18<r_{kron}<19$ sample this results in 507,813 ($\sim350$ deg$^{-2}$) and 839,208 ($\sim320$ deg$^{-2}$) galaxies in the NGC and SGC respectively, with the $19<r_{kron}<20$ sample containing 1,567,450 ($\sim1100$ deg$^{-2}$) galaxies in the NGC and 2,589,744 ($\sim1000$ deg$^{-2}$) galaxies in the SGC. For the $19<r_{kron}<20$ sample, we find 6,072,488 ($\sim2,314$ deg$^{-2}$) and 3,522,801 ($\sim2,426$ deg$^{-2}$) galaxies in the NGC and SGC. The mean redshift of our 18-19, 19-20 and 20-21 magnitude limited samples are $\bar{z}\approx0.20\pm0.09, 0.27\pm0.13$ and $0.36\pm0.16$ respectively.  

\subsection{BOSS DR12 LOWZ, CMASS and eBOSS DR14 LRG samples}
\label{sec:BOSS_eBOSS_data}

The LOWZ samples covers an area of $\sim8337$ deg$^2$ with a number density of $\sim38$ deg$^{-2}$. As $z\lesssim0.4$ LRGs were targeted in the LOWZ sample, we remove the lower redshift objects by imposing a redshift cut of $z>0.23$, thus achieving a subset of the LOWZ sample with a mean redshift of $\bar{z}=0.35$. The CMASS sample covers an area of $\sim9376$ deg$^2$ with a number density of $\sim91$ deg$^{-2}$ and an effective redshift of $z\approx0.57$. A full description of the target selection criteria for these samples is provided by \cite{Reid2016}. 

The eBOSS LRG target selection is fully described in \cite{Prakash2016}, with the sample used here containing $\sim141,000$ LRGs, covering an area of $\sim1670$ deg$^2$, resulting in an LRG number density of $\sim84$ deg$^{-2}$ with a median redshift of $z\approx0.7$. A summary of the above information for our BOSS and eBOSS LRG samples is provided in Table~\ref{tab:LRG_samples}.

\section{Methodology}
\label{sec:Methodology}

\subsection{Measuring LRG angular auto-correlation function}
\label{sec:ACF Method}

We measure the angular correlation function $\omega(\theta)$ of our LRG samples using the Landy-Szalay estimator (\citealt{LS1993}):
\begin{equation}
\omega(\theta)=1+\left(\frac{N_r}{N_d}\right)^2\frac{DD(\theta)}{RR(\theta)}-2\left(\frac{N_r}{N_d}\right)\frac{DR(\theta)}{RR(\theta)},
\label{eq:LS}
\end{equation} 
where $DD(\theta)$, $DR(\theta)$ and $RR(\theta)$ are data-data, data-random, and random-random pair counts at an angular separation of $\theta$. We perform this calculation using the CUTE\footnote{https://github.com/damonge/CUTE} algorithm (\citealt{Alonso2012}). The correlation function is calculated up to $\theta=100$ arcmin (using 19 logarithmically spaced bins), to match the range covered by \cite{Sawangwit2011} and allow for the comparison of the two results. For each sample we generate random catalogues with $20\times$ the mean number density of LRGs in the NGC and SGC and apply the same masks as applied to the data.

In order to obtain an estimate of the errors on the correlation functions we divide each sample into $N_s=6$ non-overlapping subsamples (with 2 in the NGC and 4 in the SGC), each $\sim668$ deg$^2$ in area. The mean number of LRGs in each subsample are $\sim15,800$, $\sim41,800$ and $\sim33,600$ for our $\bar{z}=0.35, 0.55$ and $0.68$ samples respectively. We then calculate the mean of these measurements, $\bar{\omega}(\theta)$, for each sample and simply take the standard error on the mean $\sigma_{\bar{\omega}(\theta)}$, as the uncertainty on the correlation function:
\begin{equation}
\sigma_{\bar{\omega}(\theta)}=\frac{\sigma_{N_s-1}}{\sqrt{N_s}}=\sqrt{\frac{\sum(\omega_i(\theta)-\bar{\omega}(\theta))^2}{N^2_s-N_s}}.
\end{equation}Here the sample standard deviation $\sigma_{N_s-1}$ is normalized to $N_s-1$ (as the mean is determined from the same dataset, reducing the number of degrees of freedom by one), and $\omega_i(\theta)$ is the correlation function of the $i$-th subsample.               
\subsection{Measuring LRG-CMB cross-correlation}
\label{sec:XCF Method}

We adopt a similar approach to \cite{Sawangwit2010} in calculating the LRG-CMB cross-correlation, a summary of which is presented here. In this work, we use the \textsc{npt} (N-point spatial statistic; \citealt{NPT2004}) code to perform the cross-correlation analysis. First \textsc{healpix} (\citealt{Gorski2005}) is used to create LRG distribution maps by dividing our LRG samples into spherical pixels of equal area, matching the resolution of our Planck CMB temperature map ($N_{side}=512$; FWHM=20 arcmin). We combine our LRG mask with the Planck CMB temperature mask and apply it to both the LRG distribution and CMB temperature maps.

The LRG distribution map is then used to calculate the LRG number over-density, $\delta_{L}(\hat{n})$, per pixel: 
\begin{equation}
\delta_{L}(\hat{n})=\frac{n_{L}(\hat{n})-\bar{n}_{L}}{\bar{n}_{L}} ,
\label{eq:delta_lrg}
\end{equation} where $n_{L}$ is the number of LRGs in a given pixel and $\bar{n}_{L}$ is the mean number of LRGs for the sample being studied.

In the case of CMASS and eBOSS spectroscopic samples, we include the associated weights when calculating the LRG over-density:
\begin{equation}
\delta_{L}(\hat{n})=\frac{n_{L}(\hat{n})-w_{tot}\times\bar{n}_{L}}{w_{tot}\times\bar{n}_{L}},
\label{eq:delta_lrg_weights}
\end{equation} where $w_{tot}=w_{systot}\times(w_{cp}+w_{noz}-1)$. Here $w_{systot}=w_{see}\times w_{star}$, is the  angular systematic weight, introduced to account for non-cosmological fluctuations in target density with stellar density and seeing, $w_{cp}$ accounts for fibre collisions and $w_{noz}$ corrects for redshift failures by up-weighting the nearest neighbour. A more detailed description of these weights is presented by \cite{Ross2012}. We do not include any weights when measuring the ISW amplitude using the LOWZ sample, as systematic weights were not supplied with the DR12 LOWZ catalogue. As inclusion of weights do not appear to have a significant impact on our CMASS and eBOSS ISW measurements however, the impact of weights on our LOWZ ISW measurement is also likely to be small.

We then calculate the LRG-CMB two-point angular cross-correlation function, $\omega_{LC}(\theta)$, using: 
\begin{equation}
\omega_{LC}(\theta)=\frac{\sum_{ij}f_{i}\delta_{L}(\bar{n_i})f_j\Delta_T(\hat{n}_j)}{\sum_{ij}f_{i}f_{j}},
\label{eq:XCF}
\end{equation}with $f_i$ representing the fraction of the $i$-th pixel located within the unmasked area, $\hat{n_i}.\hat{n_j}=cos(\theta)$ and $\Delta_T$ being the Planck CMB temperature anisotropy after removing the monopole and dipole contribution. As we are using a high pixel resolution however, the contribution from the factors weighting for unmasked fractions become negligible, and we simplify equation~(\ref{eq:XCF}) to $\omega_{LC}(\theta)=\langle\delta_{L}(\hat{n_1})\Delta_T(\hat{n}_2)\rangle$. Here we measure the cross-correlation function using 14 logarithmically spaced bins covering the range of $\theta<1400$ arcmin.    

In order to account for the correlation between the bins in the correlation function and obtain an accurate estimation of the significance of the results, we have to consider the full covariance matrix $C_{ij}$ when fitting a model to the data. Ideally, the covariance matrix is calculated based on thousands of simulated mock catalogues. However, creating such mock catalogues is a complex and computationally extensive task which lies beyond the immediate scope of this work. As a result, here we follow the technique used by \cite{Sawangwit2010} and obtain the covariance matrix using the jackknife re-sampling technique, dividing the masked Planck CMB temperature and ATLAS LRG over-density maps into 36 fields of equal area (24 in SGC and 12 in NGC). Based on these $N_{JK}=36$ jackknife subsamples are generated, omitting one field at a time. The covariance matrix is then given by:
\begin{equation}
\begin{split}
C_{ij}=&\frac{N_{JK}-1}{N_{JK}}\sum\limits_{n=1}^{N_{JK}}[(\omega_{LC,n}(\theta_i)-\bar{\omega}_{LC}(\theta_i))\times\\
&((\omega_{LC,n}(\theta_j)-\bar{\omega}_{LC}(\theta_j))],
\label{cov_mat}
\end{split}
\end{equation} where $\omega_{LC,n}(\theta_i)$ is the measured cross-correlation  of the $n$-th subsample, $\bar{\omega}_{LC}(\theta_i)$ is the mean of the measurements from all subsamples and $i$ and $j$ denote the $i$-th and $j$-th bins. The $N_{JK}-1$ factor is required in order to account for the fact that the subsamples are not independent and the uncertainty on each angular bin of the cross-correlation function $\sigma_{\omega_{LC}}(\theta)$, is given by the square root of the diagonal elements of the covariance matrix. 

For each of our samples, we obtain separate measurements of $\omega_{LC}(\theta)$ in the NGC and SGC which are combined by taking the weighted mean $\hat{\omega}_{LC}(\theta)$, of the two measurements:
\begin{gather}
\hat{\omega}_{LC}(\theta)=\frac{\sum_m\omega_{LC,m}(\theta)/\sigma_{\omega_{LC,m}(\theta)}^2}{\sum_m1/\sigma_{\omega_{LC,m}(\theta)}^2},
\end{gather}where $m$ denotes the measurement from NGC/SGC and the error on the weighted mean $\sigma_{\hat{\omega}_{LC}(\theta)}=\sqrt{1/\sum_m1/\sigma_{\omega_{LC,m}(\theta)}^2}$.

Given that our samples cover the same range of redshifts as those of \cite{Sawangwit2010}, and we have assumed the same fiducial cosmology, in this work we do not generate independent theoretical predictions for the ISW signal. Instead we simply compare our results with the models calculated in Section 3 of \cite{Sawangwit2010} based on $\Lambda$CDM predictions. 

Using the covariance matrix, we can then calculated the $\chi^2$ parameter providing a statistical measure of the quality of the fit provided by the model to our observations. The $\chi^2$ is given by: 
\begin{multline}
\chi^2=[\hat{\omega}_{LC,obs}(\theta)-{\omega}_{LC,mod}(\theta)]^TC^{-1}\\
[\hat{\omega}_{LC,obs}(\theta)-{\omega}_{LC,mod}(\theta)],
\label{eq:chi2}
\end{multline}where $\hat{\omega}_{LC,obs}(\theta)$ is our measured cross-correlation and ${\omega}_{LC,mod}(\theta)$ is the prediction from the model\footnote{See section~\ref{sec:ATLAS_SDSS_XFC} for a discussion of why we ultimately adopt an alternative approach to $\chi^2$, in accessing the level of agreement between our results and the model.}.

\section{Results and Discussion}
\label{sec:Results}

\subsection{VST ATLAS LRG angular auto-correlation function}
\label{sec:ACF}

\begin{figure*}
	\includegraphics[width=0.75\textwidth]{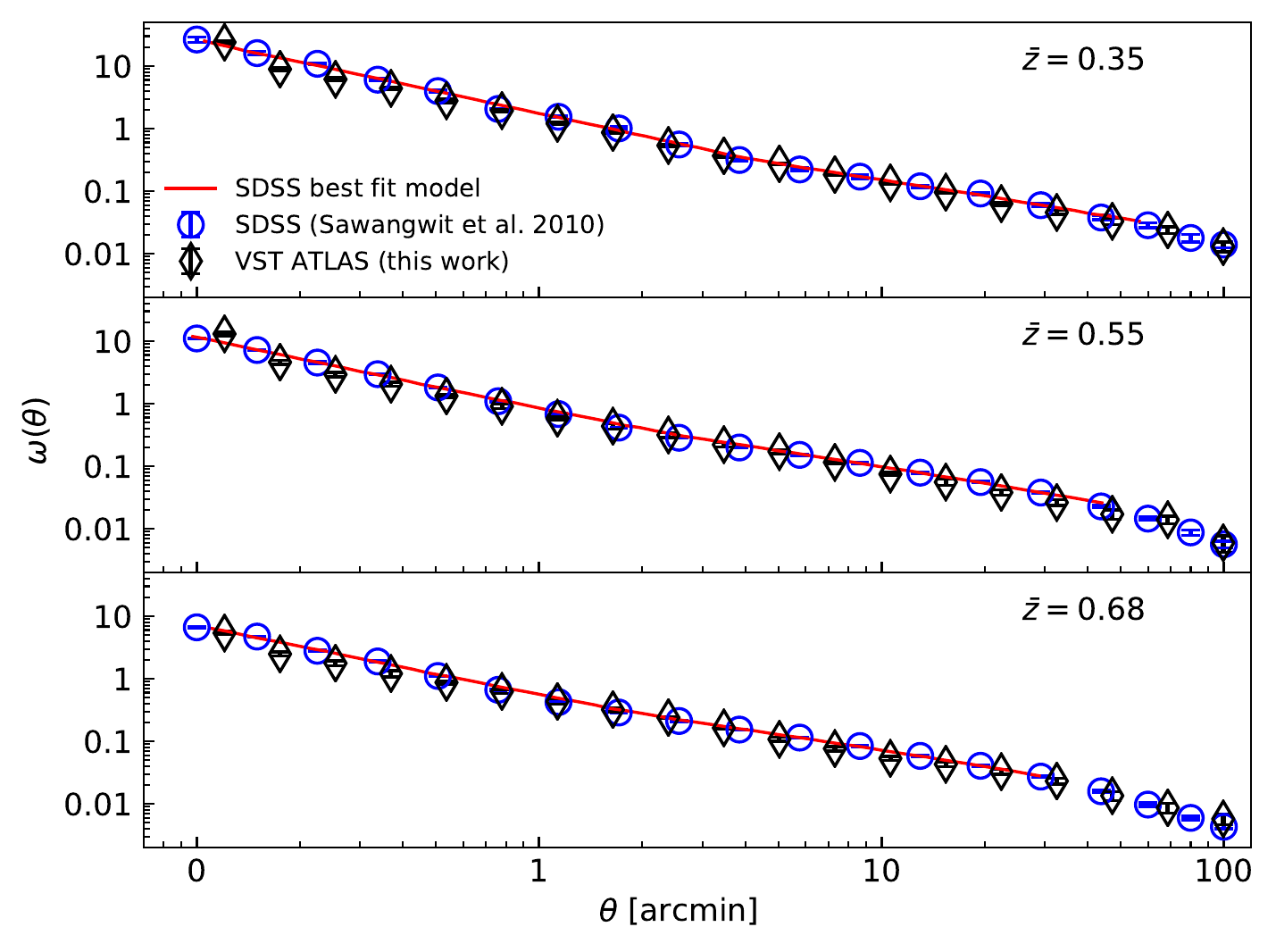}
	\caption{The VST ATLAS LRG angular auto-correlation functions, for our low, intermediate, and high redshift samples  (diamonds). The SDSS measurements of \citet{Sawangwit2011} (circles) and their best-fit double power-law models (solid lines) are added for comparison. Here, the error bars are shown inside the open data points. The good agreement between the measurements from the two datasets is an indication of the success of our LRG photometric selection in limiting the samples to the correct redshift range as well as efficient removal of stellar contamination.}
	\label{fig:ACF} 
\end{figure*}

The angular auto-correlation functions for our low, intermediate and high redshift LRG samples are presented in Figure~\ref{fig:ACF}. For all three samples we find a reasonable agreement between our results and the SDSS measurements of \cite{Sawangwit2011}. In all cases the agreement between the auto-correlation function amplitudes of the ATLAS and SDSS LRGs (and best-fit double power law models), is an indication of the success of our applied photometric selection criteria at extracting similar LRG samples from the VST ATLAS survey as those extracted from SDSS. Given the sensitivity of the auto-correlation function amplitude to stellar contamination, these results also show that our cuts have succeeded in efficiently reducing stellar contamination in our three LRG samples. Furthermore, even though our LRG samples have different number densities compared to those of \cite{Sawangwit2010}, the agreement between the ATLAS and SDSS auto-correlation functions suggests that the LRG clustering amplitude is preserved in our samples. As a result, we do not expect our measurements of the ISW amplitude to be influenced by our different sample number densities. We believe our lower LRG densities are in part due to the slightly larger scatter in the VST ATLAS colours used in the LRG sample selections, compared to the colour scatter in SDSS. Another factor influencing our lower number density could be our additional Aperture vs Kron magnitude cuts applied to remove residual stellar contamination as described in section~\ref{sec:LRG_data}.

Further tests of impact of survey systematics due to excess stellar contamination, galactic dust extinction and variations in airmass and seeing are presented in Appendix \ref{sec:Appendix_B}. Our tests indicate that these systematics do not have a significant effect on our ISW measurements.

\subsection{VST ATLAS LRG-Planck CMB cross-correlation}
\label{sec:ATLAS_SDSS_XFC}

\begin{figure*}
	\begin{subfigure}[t]{\columnwidth}
		\includegraphics[width=\columnwidth]{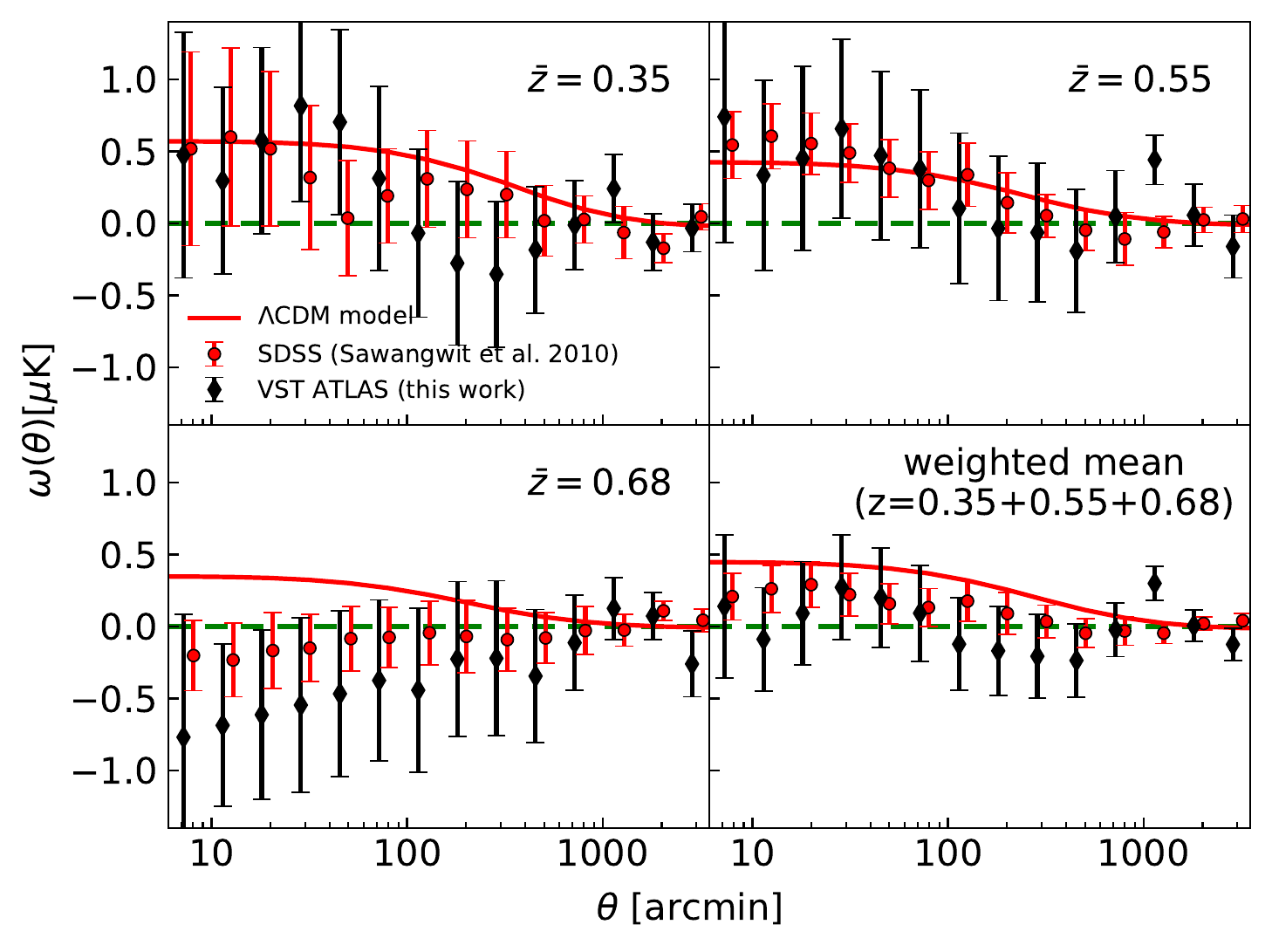} 
		\caption{VST ATLAS versus SDSS.}
		\label{fig:XCF}
	\end{subfigure}
	\vspace{0.25cm}
	\begin{subfigure}[t]{\columnwidth}
		\includegraphics[width=\columnwidth]{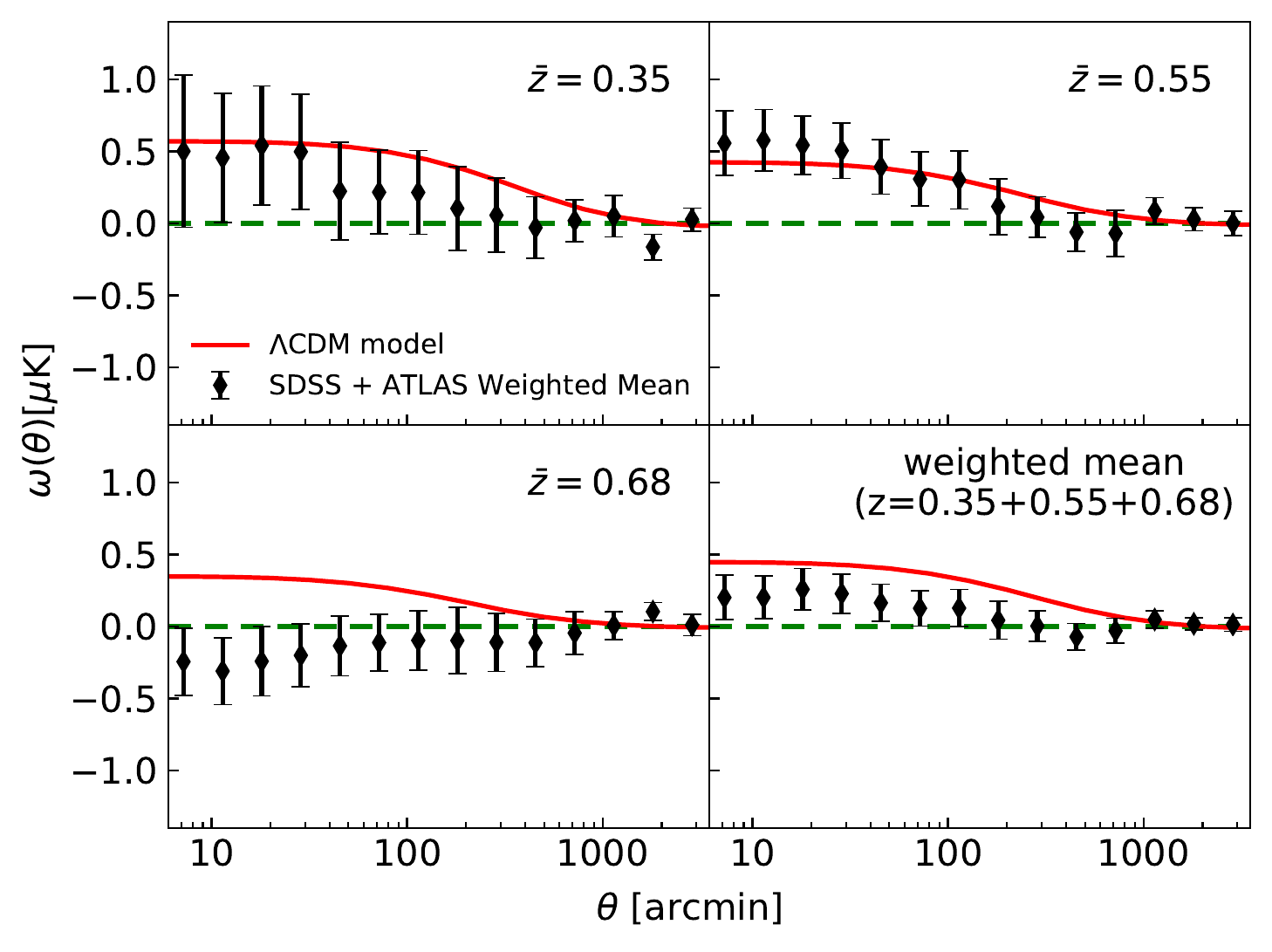} 
		\caption{VST ATLAS + SDSS.}
		\label{fig:XCF_combined} 
	\end{subfigure}
	\caption[]{(a) The VST ATLAS LRG-Planck CMB cross-correlation signal from our low, intermediate and high redshift samples compared to SDSS LRG-WMAP CMB measurements of of \cite{Sawangwit2010}. The predictions of the $\Lambda$CDM model are shown by the red solid lines. (b) The weighted mean of the two measurements in (a).}
	\label{fig:XCF_plots}
\end{figure*}

Figure~\ref{fig:XCF} shows a comparison of our ISW measurements based on the cross-correlation of VST ATLAS LRGs and Planck CMB temperature anisotropy map, to the results of \cite{Sawangwit2010} (where the same analysis was perform using SDSS LRGs and the WMAP temperature map). We find a good agreement between the two measurements in terms of ISW amplitude at all redshifts. Our error bars are however larger than those of \cite{Sawangwit2010}, which can be partially attributed to the $\sim2\times$ lower sky coverage of the ATLAS survey compared to SDSS, as well as the lower number density of LRGs, at least in the case of our $\bar{z}=0.55$ and $\bar{z}=0.68$ samples (see Table~\ref{tab:LRG_samples}). 

\begin{figure*}
	\begin{subfigure}[t]{\columnwidth}
		\includegraphics[width=\columnwidth]{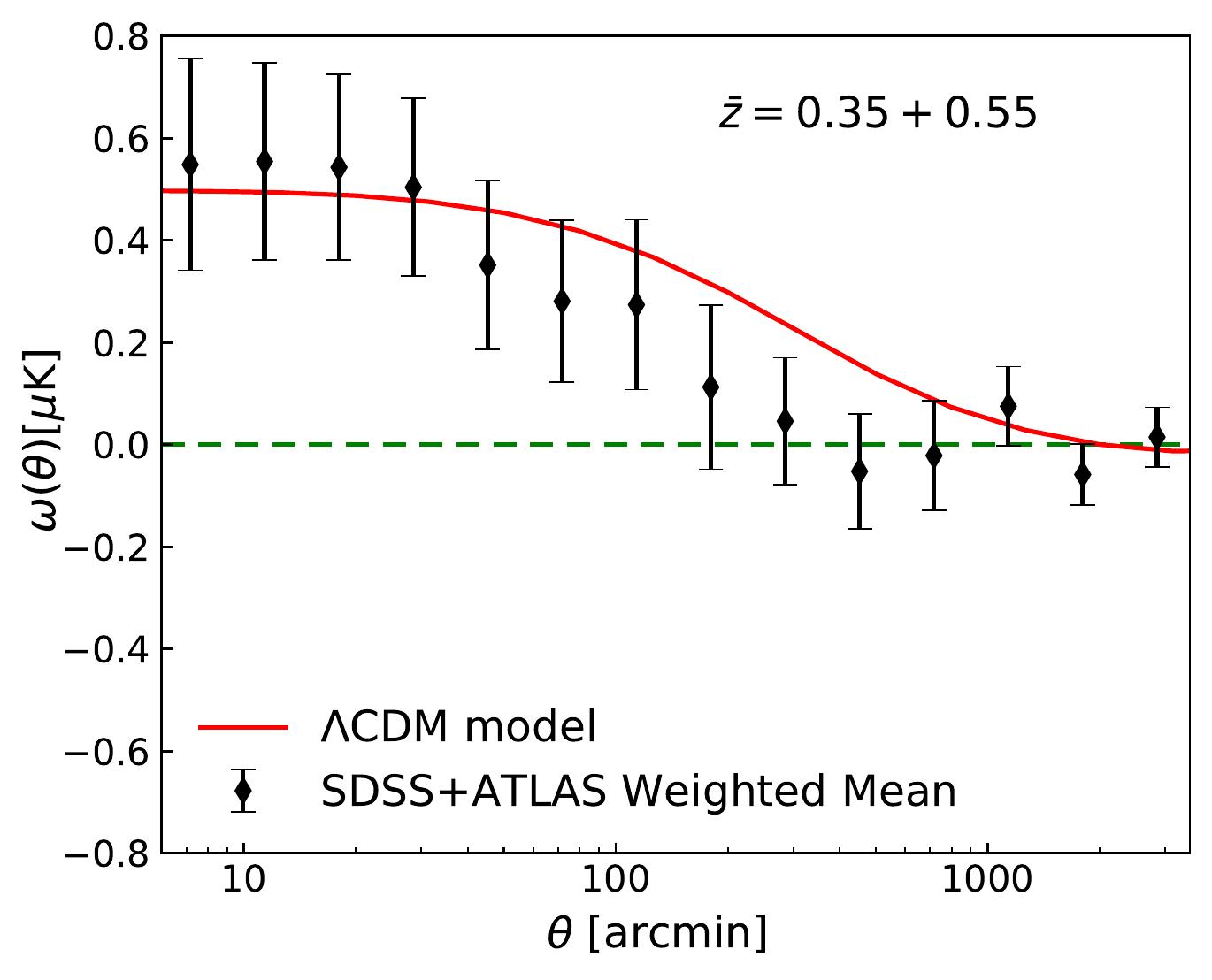} 
		\caption{}
		\label{fig:2_panel_a} 
	\end{subfigure}
	\begin{subfigure}[t]{\columnwidth}
		\includegraphics[width=\columnwidth]{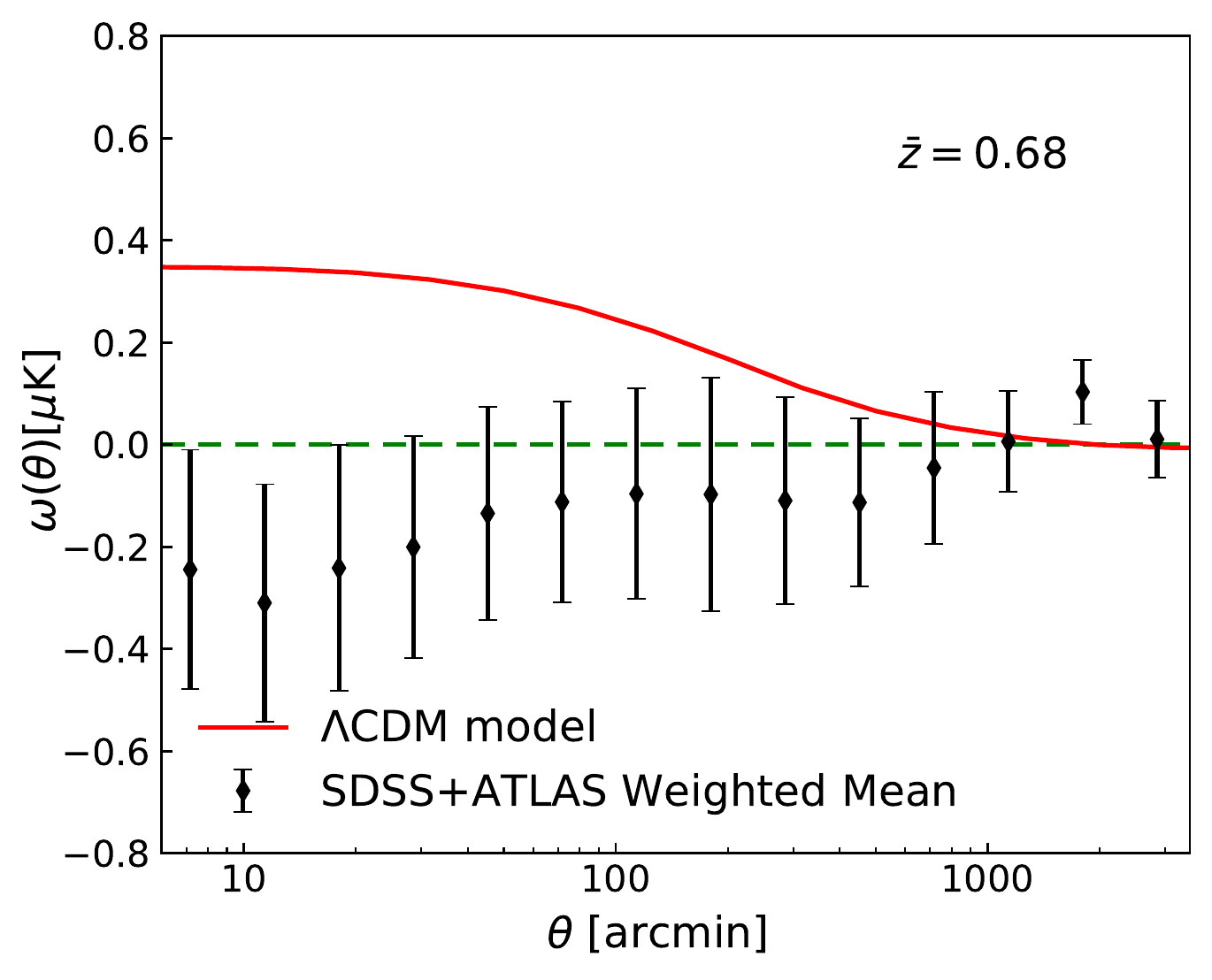} 
		\caption{}
		\label{fig:2_panel_b}
	\end{subfigure}
	\caption[]{(a) The weighted mean of the SDSS+ATLAS ISW measurements at $\bar{z}=0.35$ and $0.55$ (i.e the top panels of Figure~\ref{fig:XCF_combined}). (b) The $\bar{z}=0.68$ measurement (bottom left panel of Figure~\ref{fig:XCF_combined}). As shown in Table~\ref{tab:fitting_results}, when combining the $\bar{z}=0.35$ and $0.55$ measurements, we detect the ISW signal at $2.6\sigma$ and we note the result is in agreement with the predictions of $\Lambda$CDM. However, at $\bar{z}=0.68$ the ISW amplitude is close to zero and deviates from the $\Lambda$CDM predictions by $2.0\sigma$. }
	\label{fig:XCF_plots}
\end{figure*}

\begin{table*}
	\centering
	\caption{Summary of our VST ATLAS and BOSS/eBOSS LRGs-Planck CMB cross-correlation measurements of the ISW amplitude (based on a single bin covering the $12<\theta<120$ arcmin range). The SDSS LRG-WMAP CMB measurements of \citet{Sawangwit2010} and weighted mean of the results from the various datasets are also included. The final column shows the deviation of each measurement from the predictions of the $\Lambda$CDM model and a null amplitude. In all cases, we use the weighted mean to combine the results from different redshifts.}
	\label{tab:fitting_results}
	\begin{tabular}{c|c|c|c} % four columns, alignment for each
		\hline
		Sample & $\bar{z}$ & $\omega$(12-120 arcmin) & Deviation significance \\
	    & & [$\mu$K] & ($\Lambda$CDM, null)   \\
		\hline
		  & 0.35 & $0.47\pm0.62$ & ($0.0\sigma$, $0.8\sigma$) \\
		 VST ATLAS  & 0.55  & $0.41\pm0.51$ & ($0.1\sigma$, $0.8\sigma$) \\
		 (this work)& 0.68 & $-0.49\pm0.59$ & ($1.3\sigma$, $0.8\sigma$) \\
		 & 0.35+0.55+0.68 & $0.11\pm0.33$ & ($0.8\sigma$, $0.3\sigma$)\\
		\hline
		 & 0.35 & $0.33\pm0.33$ & ($0.5\sigma$, $1.0\sigma$)  \\
		SDSS  & 0.55 & $0.44\pm0.21$ & ($0.5\sigma$, $2.1\sigma$) \\
		\citep{Sawangwit2010}& 0.68 & $-0.13\pm0.20$ & ($2.0\sigma$, $0.6\sigma$) \\
		& 0.35+0.55+0.68 & $0.21\pm0.14$ & ($1.2\sigma$, $1.5\sigma$) \\	
		\hline
			 & 0.35 & $0.32\pm0.38$ & ($0.4\sigma$, $0.9\sigma$)  \\
		BOSS/eBOSS  & 0.55 & $0.73\pm0.38$ & ($1.1\sigma$, $2.0\sigma$) \\
		(this work) & 0.68 & $0.50\pm0.76$ & ($0.3\sigma$, $0.7\sigma$) \\
		& 0.35+0.55+0.68 & $0.52\pm0.25$ & ($0.6\sigma$, $2.1\sigma$) \\	
		\hline
		\hline
		 & 0.35 & $0.36\pm0.29$ & ($0.5\sigma$, $1.2\sigma$)  \\
		VST ATLAS+SDSS  & 0.55 & $0.43\pm0.19$ & ($0.5\sigma$, $2.3\sigma$) \\
		(weighted mean) & 0.68 & $-0.17\pm0.19$ & ($2.3\sigma$, $0.9\sigma$) \\
		& 0.35+0.55 & $0.41\pm0.16$& ($0.1\sigma$, $2.6\sigma$) \\
		& 0.35+0.55+0.68 & $0.20\pm0.12$& ($1.4\sigma$, $1.7\sigma$) \\
				\hline
		& 0.35 & $0.36\pm0.32$ & ($0.4\sigma$, $1.1\sigma$)  \\
VST ATLAS+BOSS/eBOSS & 0.55 & $0.62\pm0.31$ & ($0.9\sigma$, $2.0\sigma$) \\
		(weighted mean) & 0.68 & $-0.12\pm0.47$ & ($0.8\sigma$, $0.3\sigma$) \\
		& 0.35+0.55 & $0.49\pm0.22$ & ($0.4\sigma$, $2.2\sigma$) \\
		& 0.35+0.55+0.68 & $0.38\pm0.20$ & ($0.1\sigma$, $1.9\sigma$) \\
	\end{tabular}
\end{table*}

As on large scales relevant to ISW measurements the statistical error is limited by sample variance, one would expect the errors on our ISW measurements to scale with $\sigma_{ATLAS}/\sigma_{SDSS}\approx\sqrt{(A_{SDSS}/A_{ATLAS})}$ where $A$ represents the area of each sample. We therefore expect the ATLAS error bars to be $\sim1.4\times$ larger than those of SDSS. However, we find that the errors on our VST ATLAS LRG ISW measurements do not obey the above scaling with SDSS and are $\sim1.9\times$, $\sim2.4\times$ and $\sim3.0\times$ larger than those from SDSS, for our $\bar{z}=0.35$, $0.55$ and $0.68$ samples respectively. Assuming the SDSS ISW errors of \cite{Sawangwit2010} are not under-estimated, the reason behind the larger than expected errors on our LRG ISW measurements remains unknown.

\begin{figure*}
    \begin{subfigure}[t]{0.96\textwidth}
		\includegraphics[width=\textwidth]{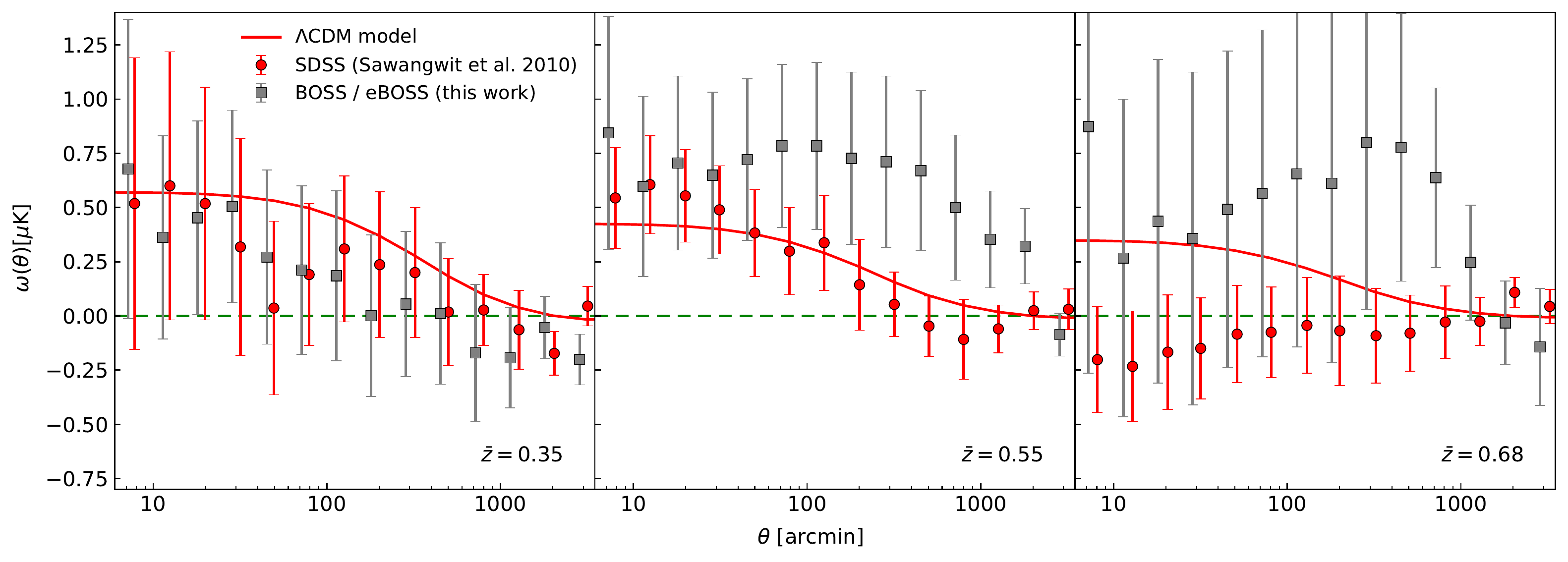} 
		\caption{SDSS versus BOSS/eBOSS.}
		\hfill
		\label{fig:BOSS_vs_SDSS} 
	\end{subfigure}
	\begin{subfigure}[t]{\columnwidth}
		\includegraphics[width=\columnwidth]{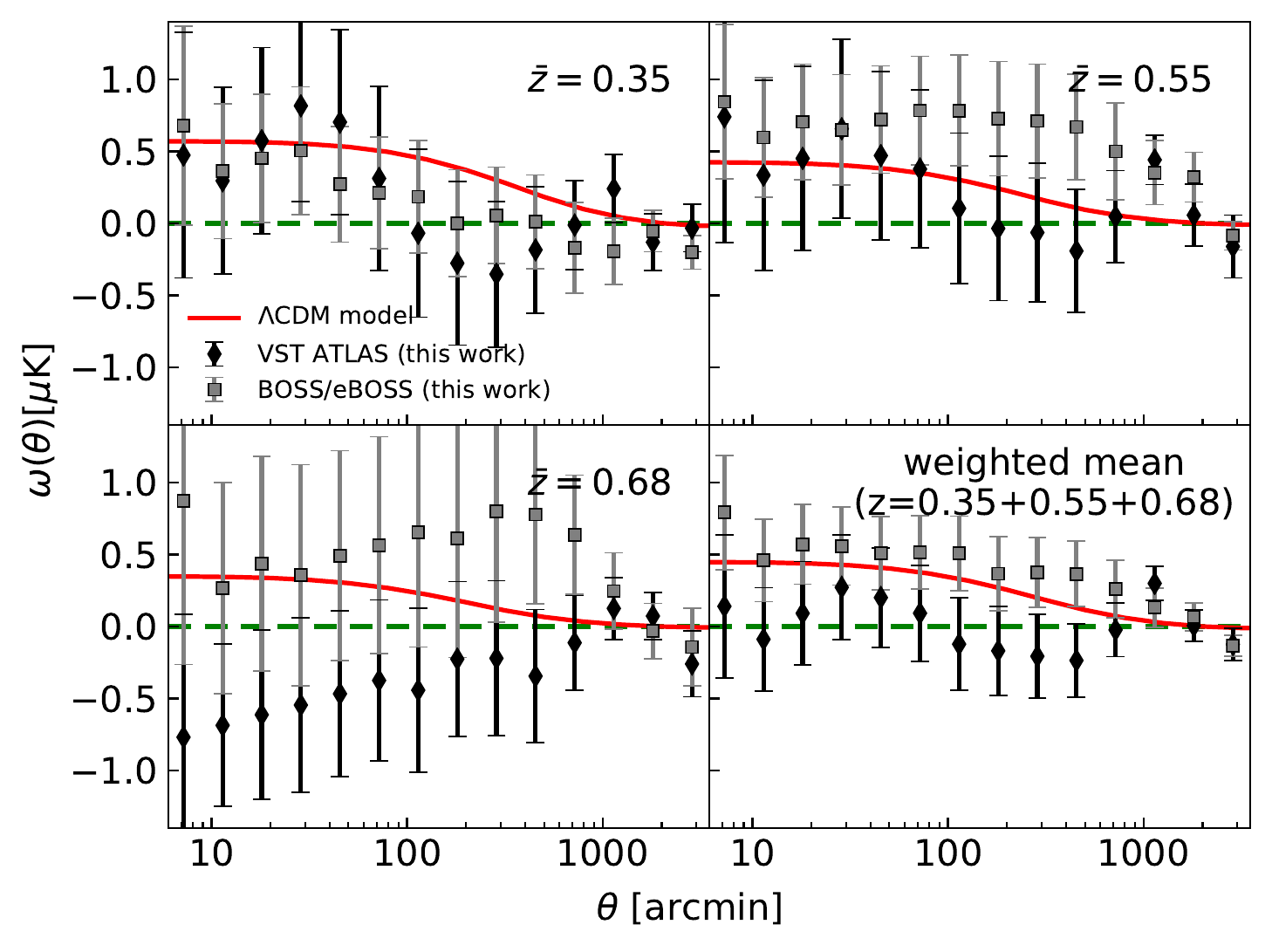} 
		\caption{VST ATLAS versus BOSS/eBOSS.}
		\label{fig:BOSS_XCF} 
	\end{subfigure}
	\begin{subfigure}[t]{\columnwidth}
		\includegraphics[width=\columnwidth]{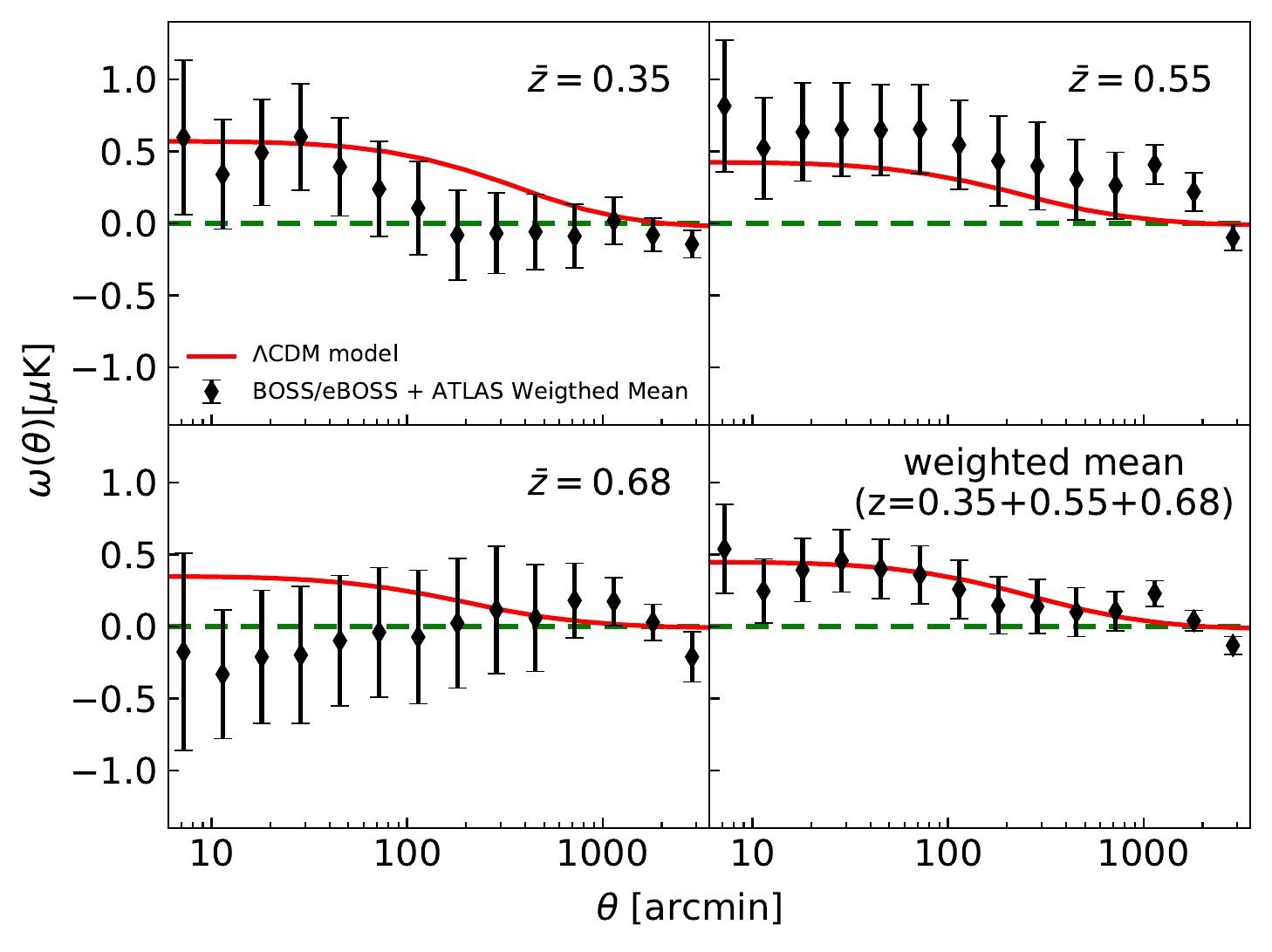}
		\caption{VST ATLAS + BOSS/eBOSS.}
		\label{fig:BOSS_XCF_combined} 
	\end{subfigure}
	\caption[]{(a) Comparison of the SDSS and BOSS/eBOSS ISW measurements. (b, c) same as (\ref{fig:XCF}, \ref{fig:XCF_combined}) but comparing / combining the VST ATLAS ISW measurements to those based on BOSS/eBOSS.} 
	\label{fig:XCF_2panel}
\end{figure*}

Similarly to \cite{Sawangwit2010}, when fitting our measurements to the $\Lambda$CDM model, we find the resulting $\chi^2$ values (given by equation~\ref{eq:chi2}) to be unreliable. This is likely due to our use of the jackknife technique in estimating the covariance matrices (see equation~\ref{cov_mat}) and the failure of this technique in accurately estimating the off-diagonal covariance matrix elements, which in turn impacts the $\chi^2$ fitting results. As mocks are currently not available for the VST ATLAS survey (and the Bootstrap technique was also unsuccessful in improving our covariance matrix estimations), we follow the approach of \cite{Sawangwit2010} and simply assess the deviation of our measurements from the $\Lambda$CDM predictions and a null ISW amplitude, based on a single large bin covering the $12<\theta<120$ arcmin range.

Table~\ref{tab:fitting_results} contains a summary of the our single bin ISW measurements, those of \cite{Sawangwit2010} and also the weighted mean of the results from the two studies (see
Figure~\ref{fig:XCF_combined}). In the case of the $\bar{z}=0.35$ and $0.55$ LRG samples we found our detected ISW amplitude to be in agreement with the predictions of $\Lambda$CDM, supporting the late-time accelerated expansion of the Universe. As seen in Table~\ref{tab:fitting_results}, upon combining the ATLAS and SDSS measurements, at these redshifts we detect the ISW effect at $1.2\sigma$ and $2.3\sigma$ (or $2.6\sigma$ combined\footnote{Based on the weighted mean of the results from the two redshifts.} - see 
Fig.~\ref{fig:2_panel_a}). In the case of the $\bar{z}=0.68$ LRG sample however, where the ISW measurement from VST ATLAS has a similar negative amplitude to SDSS, we find a $\sim2\sigma$ deviation from the $\Lambda$CDM prediction, when combining the results from the two studies (Figure~\ref{fig:2_panel_b}).  

In these measurements, the signal is mostly dominated by SDSS and combining the VST ATLAS and SDSS results only yields a small increase in the significance of detection (or rejection) of the $\Lambda$CDM ISW predictions, compared to the results previously obtained from SDSS alone. We note however, that the errors in the ATLAS ISW measurements would be 50-70\% smaller if they had scaled correctly with sample size, which may explain the unexpectedly good agreement between SDSS and ATLAS results in all three redshift ranges. Overall, the results of this study offer a valuable confirmation of the measured ISW amplitudes of \cite{Sawangwit2010} based on SDSS and WMAP, using the cross-correlation of two independent datasets (VST ATLAS and Planck) that also cover completely separate areas of the sky.  

\subsection{Comparison to BOSS DR12 LOWZ, CMASS and eBOSS LRGs samples}
\label{sec:LOWZ_CMASS_eBOSS_XFC}

To further verify the SDSS measurements at $\bar{z}=0.35$, $0.55$ and $0.68$ we compare the results with those obtained using the LOWZ, CMASS and eBOSS LRG redshift samples (see Section~\ref{sec:BOSS_eBOSS_data}). As shown in Figure~\ref{fig:BOSS_vs_SDSS} with the exception of the $\theta>100$ arcmin higher BOSS ISW amplitude at $\bar{z}=0.55$, the BOSS measurements provide a general confirmation of the SDSS results at $\bar{z}=0.35$ and $\bar{z}=0.55$. The $\bar{z}=0.35$ results show particularly good agreement between the photometric and spectroscopic samples. The reason behind the higher than expected $\bar{z}=0.55$ BOSS amplitude at large separations remains unknown. At $\bar{z}=0.68$, the  ISW amplitude is more positive in the eBOSS LRG sample than observed in SDSS or ATLAS (see Figures~\ref{fig:BOSS_vs_SDSS} and \ref{fig:BOSS_XCF}). 
Nevertheless, the eBOSS result shows a qualitatively different
form to that of the lower redshift results generally rising towards larger separations rather than falling. This behaviour is also similar to that seen in SDSS and ATLAS at the same redshift, just with a higher amplitude for eBOSS.

At $\bar{z}=0.35$ and $\bar{z}=0.55$, we therefore find similar results whether we combine ATLAS with SDSS photometric, or BOSS/eBOSS spectroscopic LRG samples. For example, in Table~\ref{tab:fitting_results}, at $\bar{z}=0.35$ and $0.55$, the null amplitude is rejected at $1.1$ and $2.0\sigma$ when combining the ATLAS+BOSS measurements (Figure~\ref{fig:BOSS_XCF_combined}); similar to the $1.2$ and $2.3\sigma$ ATLAS+SDSS ISW detection. When combining the measurements at $\bar{z}=0.35$ and $0.55$, the ATLAS+BOSS result rejects the null signal at $2.2\sigma$, compared to the $2.6\sigma$ null rejection obtained from ATLAS+SDSS. 

At $\bar{z}=0.68$ however, Table~\ref{tab:fitting_results} shows a $0.8\sigma$ deviation from $\Lambda$CDM rather than $2.3\sigma$, when the ATLAS measurement is combined with eBOSS instead of SDSS. Similarly, the ATLAS+BOSS/eBOSS weighted mean of the results from the 3 redshift bins appears to be in better agreement with $\Lambda$CDM compared to ATLAS+SDSS (a $0.1\sigma$ deviation compared to $1.4\sigma$). However, in both cases this lower rejection significance of $\Lambda$CDM is mainly due to the larger eBOSS errors, rather than any intrinsically improved agreement of the {\it form} of the high redshift result to the ISW model.

These larger errors on the eBOSS ISW measurements are due to its lower sky coverage than that of the equivalent SDSS LRG sample, and SDSS thus remains the $\bar{z}=0.68$ measurement with the highest signal in this sky area. We therefore conclude that ATLAS+SDSS measurement shown in Figure~\ref{fig:2_panel_b}  provides the best estimate of the ISW effect using $\bar{z}=0.68$ LRGs, in the full North+South combined sample. Similarly, in Figure~\ref{fig:2_panel_a} we use the ATLAS and SDSS data to provide the best $\bar{z}=0.35$ plus $\bar{z}=0.55$ ISW measurement in the full North+South sample. The difference between the two appears clear, although the $\bar{z}=0.68$ deviation significance from $\Lambda$CDM, is currently only at a moderate level of $\sim2.3\sigma$. It is therefore important to re-measure the high redshift ISW signal using the complete eBOSS survey, as well as future surveys such as DESI \citep{DESI2016} and LSST \citep{LSST2019}, which will offer large, high-purity LRG samples that could assess any potential deviations from $\Lambda$CDM at a higher statistical significance.

\subsection{Magnitude limited samples}
\label{sec:Maglim_results}

\begin{figure*}
    \begin{subfigure}[t]{\textwidth}
	\includegraphics[width=\textwidth]{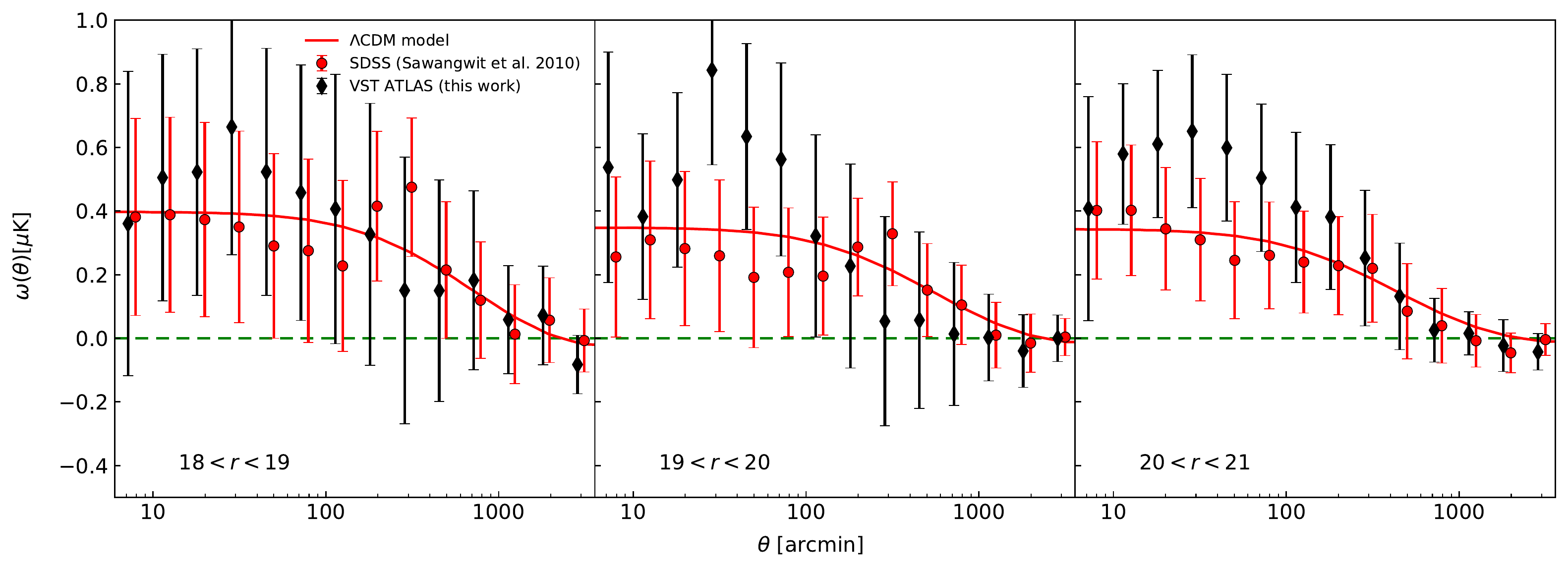}
	\caption{VST ATLAS versus SDSS.}
	\label{fig:XCF_maglim_ATLAS_SDSS}
	\end{subfigure}
	\begin{subfigure}[t]{\textwidth}
	\includegraphics[width=\textwidth]{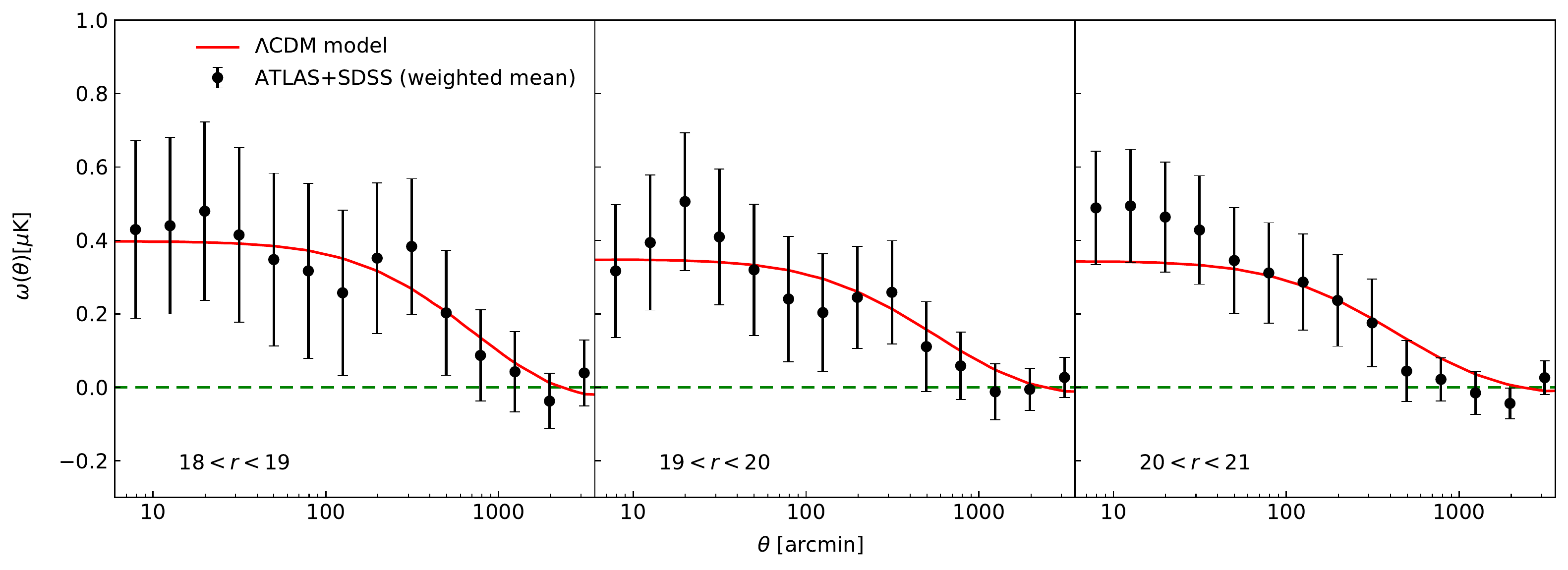}
	\caption{VST ATLAS + SDSS.}
	\label{fig:XCF_maglim_ATLAS+SDSS}
	\end{subfigure}
	\caption{(a) The ISW signal from our three VST ATLAS r-band magnitude limited galaxy samples (diamonds) in comparison with the SDSS results of \protect\cite{Sawangwit2010}  (circles) and $\Lambda$CDM model prediction (solid lines). (b) weighted mean of the results from the two studies. Here the mean redshifts of the $18<r<19$, $19<r<20$ and $20<r<21$ samples are $\bar{z}\approx0.20, 0.27$ and $0.36$ respectively.}
	\label{fig:XCF_maglim} 
\end{figure*}

Figure~\ref{fig:XCF_maglim_ATLAS_SDSS} shows a comparison of our measurements of the three $r$-band magnitude limited samples to the SDSS measurements of \cite{Sawangwit2010}. Once again a general agreement is found between the two sets of measurements. Unlike our redshift limited LRG samples, here the number of galaxies in our three samples are in line with theoretical expectations, and we find the VST ATLAS error bars to be comparable to those of \cite{Sawangwit2010} based on SDSS, once the difference in survey areas is accounted for.

Upon combining the two sets of measurements by taking their weighted mean (see Figure~\ref{fig:XCF_maglim_ATLAS+SDSS}), we find that on scales of $12<\theta<120$ arcmin, the null amplitude is rejected at moderate levels of $\sim1.3\sigma$, $\sim1.9\sigma$ and $\sim2.0\sigma$ for the $18<r<19$, $19<r<20$ and $20<r<21$ samples respectively. Recalling that these samples have mean redshifts of $\bar{z}\approx0.20\pm0.09, 0.27\pm0.13$ and $0.36\pm0.16$, we note that the $\sim2.0\sigma$ ISW detection obtained from the $20<r<21$ ATLAS+SDSS galaxy samples, provides a further confirmation of our $1.2\sigma$ ISW detection based on the $\bar{z}=0.35$ ATLAS+SDSS LRG samples. 

\subsection{ISW rotation test}
\label{sec:rotation_test}

Following the approach of previous works including \cite{Sawangwit2010} and \cite{Giannantonio2012}, we test for presence of systematic effects and the robustness of our measurements by rotating the LRG data with respect to the CMB map in increments of $40\degree$ about the Galactic pole (by adding $40\degree$ to the Galactic longitude). Here we perform the rotation test on the LOWZ and CMASS samples, as they provide contamination-free samples of spectroscopically confirmed LRGs. Given the current low sky coverage and large uncertainties on the eBOSS measurement, we do not include this sample in our rotation tests. 

\begin{figure*}
	\begin{subfigure}{\columnwidth}
	    \includegraphics[width=\columnwidth]{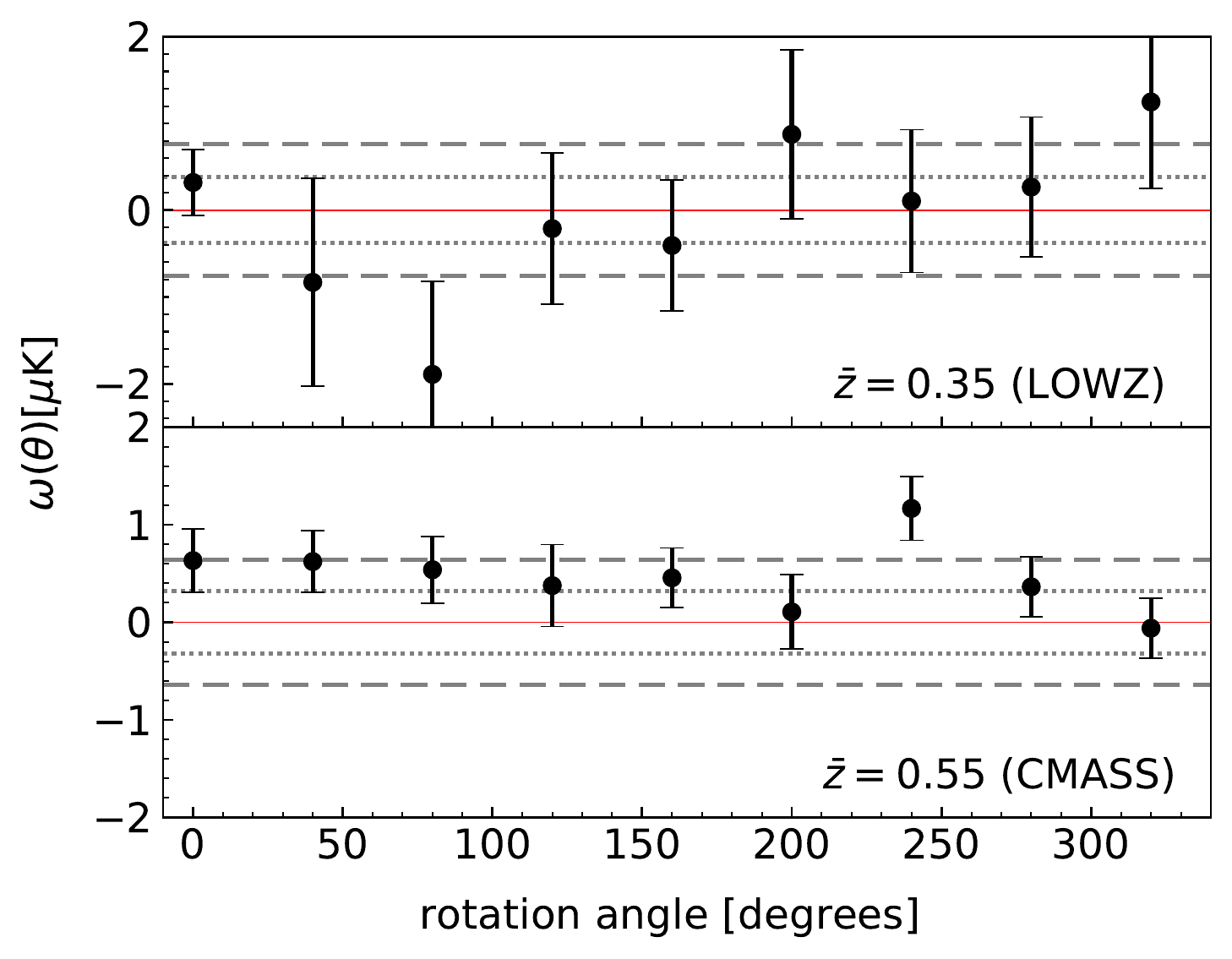}
	    \caption{}
		\label{fig:BOSS_rotation}
	\end{subfigure}
	\begin{subfigure}{\columnwidth}
		\includegraphics[width=\columnwidth]{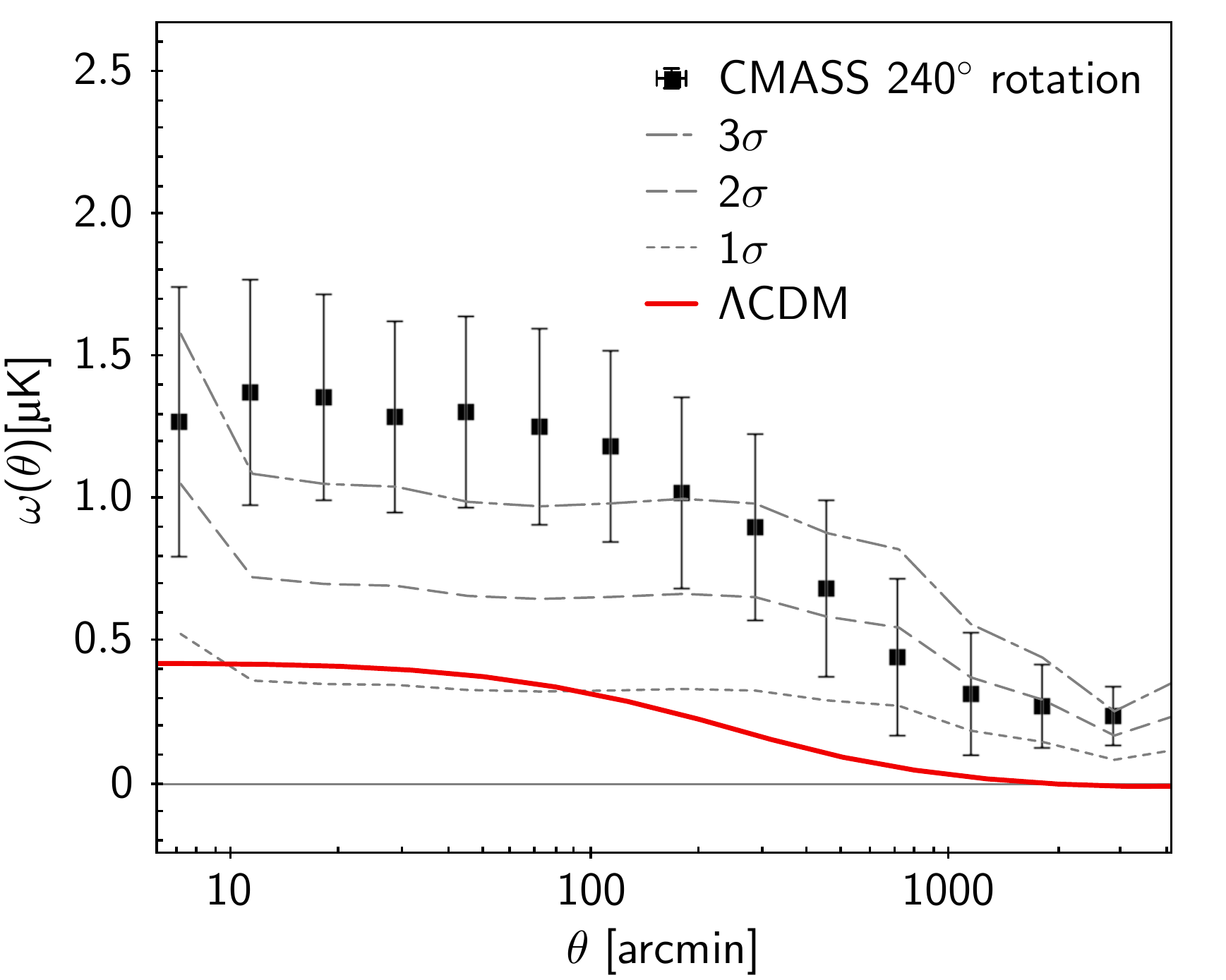} 
		\caption{}
		\label{fig:CMASS_full_rotation} 
	\end{subfigure}
	\caption[]{(a) The BOSS LOWZ and CMASS single-bin ($12<\theta<120$ arcmin) ISW amplitudes as a function of rotation angles (see text for details). The 1 and $2\sigma$ error regions around the null amplitude are shown by dotted and dashed lines. These are based on the error bars of the zero rotation data points and represent level of variance in the data. (b) The full-range $240\degree$ rotation result for the CMASS sample showing that in the in the $12<\theta<120$ arcmin range used to extract our single bin measurement, the ISW amplitude is consistently high and is not merely due to the presence of a few outliers.}
	\label{fig:rotation_tests}
\end{figure*}

Figure~\ref{fig:BOSS_rotation} shows the results of our single-bin ($12<\theta<120$ arcmin) rotation tests for the LOWZ and CMASS samples. Statistically, one would expect 32 and 5 per cent of the rotations to exceed the 1 and $2\sigma$ thresholds. 

In the case of the LOWZ sample we find that 4/8 and 2/8 (50 and 25 per cent) of the rotations result in amplitudes that lie above the 1 and $2\sigma$ thresholds respectively. Furthermore, we find that 2/8 (25 per cent) of the rotations produce a more significant rejection of null, than the zero-rotation value. 

For the CMASS sample we find that 4/8 and 1/8 (50 and 12.5 per cent) of the points are further than 1 and $2\sigma$ away from zero respectively, while 1/8 (12.5 per cent) of the rotations produces a more significant null rejection than our zero-rotation result. In order to ensure the single-bin measurements in Figure~\ref{fig:BOSS_rotation} do not appear to be artificially deviated from zero due to the presence of outliers in the full-range cross-correlation functions, in Figure~\ref{fig:CMASS_full_rotation} we show the full-range cross-correlation function for the $240\degree$ rotation of the CMASS sample (which resulted in the highest deviation from zero). Here, we can see that the $240\degree$ rotation appears to have produced an ISW amplitude which is consistently high in our $12<\theta<120$ arcmin range of interest, resulting in a higher null rejection when compared to the $2\sigma$ rejection obtained from the zero-rotation CMASS result as shown in Table~\ref{tab:fitting_results}.

Here our results are in agreement with those of \cite{Sawangwit2010}, who found that in 1 to 2/8 cases, the rotated maps produced a more significant null rejection than the zero-rotation result. However, we find our results to be in contrast to the findings of \cite{Giannantonio2012} (as shown in their Table 3), where in their 6 studied samples only 23 and 2 per cent of their rotations exceeded the 1 and $2\sigma$ thresholds (fully consistent with the 32 and 5 per cent statistical expectations), with none exceeding the null rejection significance of the unrotated map. Similarly, \cite{Giannantonio2012} found that across all 7 samples studied by \cite{Sawangwit2010}, only 39 and 11 per cent of rotations exceeded the 1 and $2\sigma$ thresholds. 

\cite{Giannantonio2012} suggest that the higher percentage of points exceeding the 1 and $2\sigma$ thresholds, found by \cite{Sawangwit2010} could be due to their use of the jackknife method in estimating the errors, which has been shown to produce somewhat smaller errors (see \citealt{Cabre2007}) than those obtained from simulated mocks (as used by \citealt{Giannantonio2012}). This could in part also explain the higher than expected percentages found in our rotation tests. However, given that for the LOWZ and CMASS samples, 50 percent of our rotations exceeded the $1\sigma$ threshold, 25 and 12.5 percent exceeded the $2\sigma$ threshold, and 2/8 and 1/8 rotations produced a more significant rejection of null than the unrotated map; our findings suggest that the robustness of current ISW detections is still not completely secure even at $\bar{z}=0.35$ and $\bar{z}=0.55$. Consequently, as well as any remaining statistical gains, improvements in reducing systematics on ISW measurements should still be sought in future works.        

\section{Conclusions}
\label{sec:Conclusions}

We have presented our measurements of the ISW signal in the cross-correlation of the Planck CMB temperature map with three photometrically selected LRG samples with mean redshifts of $\bar{z}=0.35$, $\bar{z}=0.55$ and $\bar{z}=0.68$, selected from the VST ATLAS survey. We then combine our measurements with those of \cite{Sawangwit2010}, where the same analysis was performed using the WMAP CMB temperature map and LRG samples selected from SDSS. 
 
Upon combining the measurements from ATLAS and SDSS, at $\bar{z}=0.35$ and $\bar{z}=0.55$, we detect the ISW signal at $1.2\sigma$ and $2.3\sigma$ respectively (i.e. a combined detection of $2.6\sigma$). This is in agreement with the predictions of $\Lambda$CDM supporting the late-time accelerated expansion of the Universe. We further verify our results at these redshifts by repeating the measurements using the BOSS DR12 LOWZ and CMASS spectroscopic LRG samples. This time upon combining the ATLAS and BOSS measurements, we detect the ISW signal at $1.1\sigma$ and $2.0\sigma$ (with a combined significance of $2.2\sigma$). Furthermore, we detect the ISW effect in 3 magnitude limited galaxy samples, with mean redshifts of $\bar{z}\approx0.20$, $0.27$ and $0.36$, at $\sim1.3, 1.9$ and $2.0\sigma$ respectively.

However, we do not detect the ISW signal at $\bar{z}=0.68$ when combining the ATLAS and SDSS results. Further tests using eBOSS LRGs at this redshift remain inconclusive due to the large uncertainties, caused by the current relatively low sky coverage of the survey. If the ISW signal is shown to be inconsistent with the predictions of $\Lambda$CDM at high redshifts, it could open the door to alternative theories such as modified gravity models. It is therefore important to repeat the $z\sim0.7$ ISW measurement upon the completion of the eBOSS survey and using data from upcoming surveys such as DESI and LSST which will provide the statistics and reduced systematics required to assess any deviations from the predictions of $\Lambda$CDM.
 
Finally, we test the robustness of our ISW measurements at $\bar{z}=0.35$ and $\bar{z}=0.55$ by rotating the LRG overdensity map with respect to the CMB temperature map in 8 increments about the Galactic pole. Here, in contrast to the findings of \cite{Giannantonio2012}, we find that a higher percentage of rotations result in amplitudes 1 and $2\sigma$ away from zero than statistically expected. Furthermore, we find that in the case of LOWZ and CMASS samples 2/8 and 1/8 rotations result in more significant rejections of the null amplitude than obtained from our unrotated maps. Consequently, our results indicate that the robustness and significance of ISW detections still warrant further examination in future works. Similarly rotation tests could serve as a useful tool for determining the level of systematics in ISW measurements obtained from future surveys. 

In summary, the results of this study provide a confirmation of previous ISW measurements from \cite{Sawangwit2010}. However, despite the visual impressions given by the cross-correlation measurements, our detections of the ISW signal at $\bar{z}=0.35$, $\bar{z}=0.55$ and in 3 magnitude limited samples remain at low to moderate levels of significance. However, previous works such as \cite{Francis2010} have demonstrated that the ISW signal could remain evasive in $\gtrsim10$ per cent of cases, even with the availability of the best possible data. Nonetheless, given the cosmological implications of any significant deviations from the predictions of $\Lambda$CDM, repeating the ISW measurement at $z\sim0.7$, where our results point to the possibility of such deviations, using the next generation of large sky surveys, remains a worthwhile and important endeavour. 

\section*{Acknowledgements}

This research made use of Astropy,\footnote{http://www.astropy.org} a community-developed core Python package for Astronomy \citep{astropy2018}, as well as TOPCAT \& STILTS\footnote{http://www.star.bris.ac.uk/~mbt/topcat/sun253/index.html} packages \citep{Taylor2005}. 

Funding for the Sloan Digital Sky Survey IV has been provided by the Alfred P. Sloan Foundation, the U.S. Department of Energy Office of Science, and the Participating Institutions. SDSS-IV acknowledges support and resources from the Center for High-Performance Computing at the University of Utah. The SDSS web site is www.sdss.org.

SDSS-IV is managed by the Astrophysical Research Consortium for the Participating Institutions of the SDSS Collaboration including the Brazilian Participation Group, the Carnegie Institution for Science, Carnegie Mellon University, the Chilean Participation Group, the French Participation Group, Harvard-Smithsonian Center for Astrophysics, Instituto de Astrof\'isica de Canarias, The Johns Hopkins University, Kavli Institute for the Physics and Mathematics of the Universe (IPMU) / University of Tokyo, the Korean Participation Group, Lawrence Berkeley National Laboratory,  Leibniz Institut f\"ur Astrophysik Potsdam (AIP), Max-Planck-Institut f\"ur Astronomie (MPIA Heidelberg), Max-Planck-Institut f\"ur Astrophysik (MPA Garching), Max-Planck-Institut f\"ur Extraterrestrische Physik (MPE), National Astronomical Observatories of China, New Mexico State University, New York University, University of Notre Dame, Observat\'ario Nacional / MCTI, The Ohio State University, Pennsylvania State University, Shanghai Astronomical Observatory, United Kingdom Participation Group, Universidad Nacional Aut\'onoma de M\'exico, University of Arizona, University of Colorado Boulder, University of Oxford, University of Portsmouth, University of Utah, University of Virginia, University of Washington, University of Wisconsin, Vanderbilt University, and Yale University.

RM,  TS  \&  NM  acknowledge  the  support  of  the Science and Technology Facilities Council (ST/L00075X/1and ST/L000541/1).

We would also like to thank Alina Talmantaite for her useful comments and discussion.

%%%%%%%%%%%%%%%%%%%%%%%%%%%%%%%%%%%%%%%%%%%%%%%%%%



%\bibliographystyle{mnras}

\bibliographystyle{mnras}
\bibliography{bibliography}

%\begin{thebibliography}{99}
%\bibitem[\protect\citeauthoryear{Xu et al.}{2012}]{Xu et al.2012}
%Xu X. et al., 2012, MNRAS 427, 2146–2167
%\end{thebibliography}

%%%%%%%%%%%%%%%%%%%%%%%%%%%%%%%%%%%%%%%%%%%%%%%%%%

%%%%%%%%%%%%%%%%% APPENDICES %%%%%%%%%%%%%%%%%%%%%

\appendix
\section{ATLAS / SDSS colour comparison}
\label{sec:Appendix_A}
\begin{figure*}
	\begin{subfigure}{0.66\columnwidth}
		\includegraphics[width=\columnwidth]{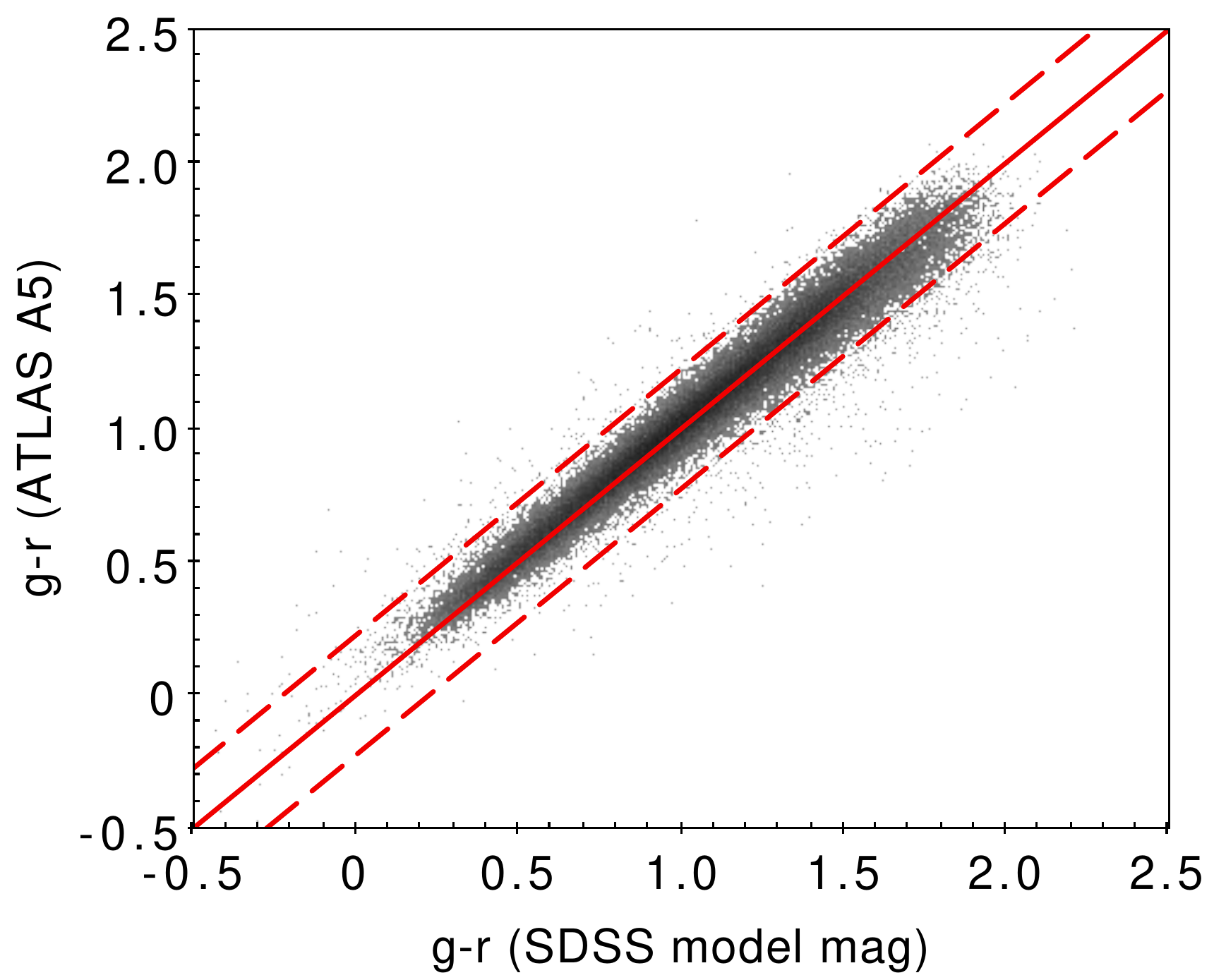} 
		\label{fig:g-r_ATLAS_SDSS}
	\end{subfigure}
		\begin{subfigure}{0.66\columnwidth}
		\includegraphics[width=\columnwidth]{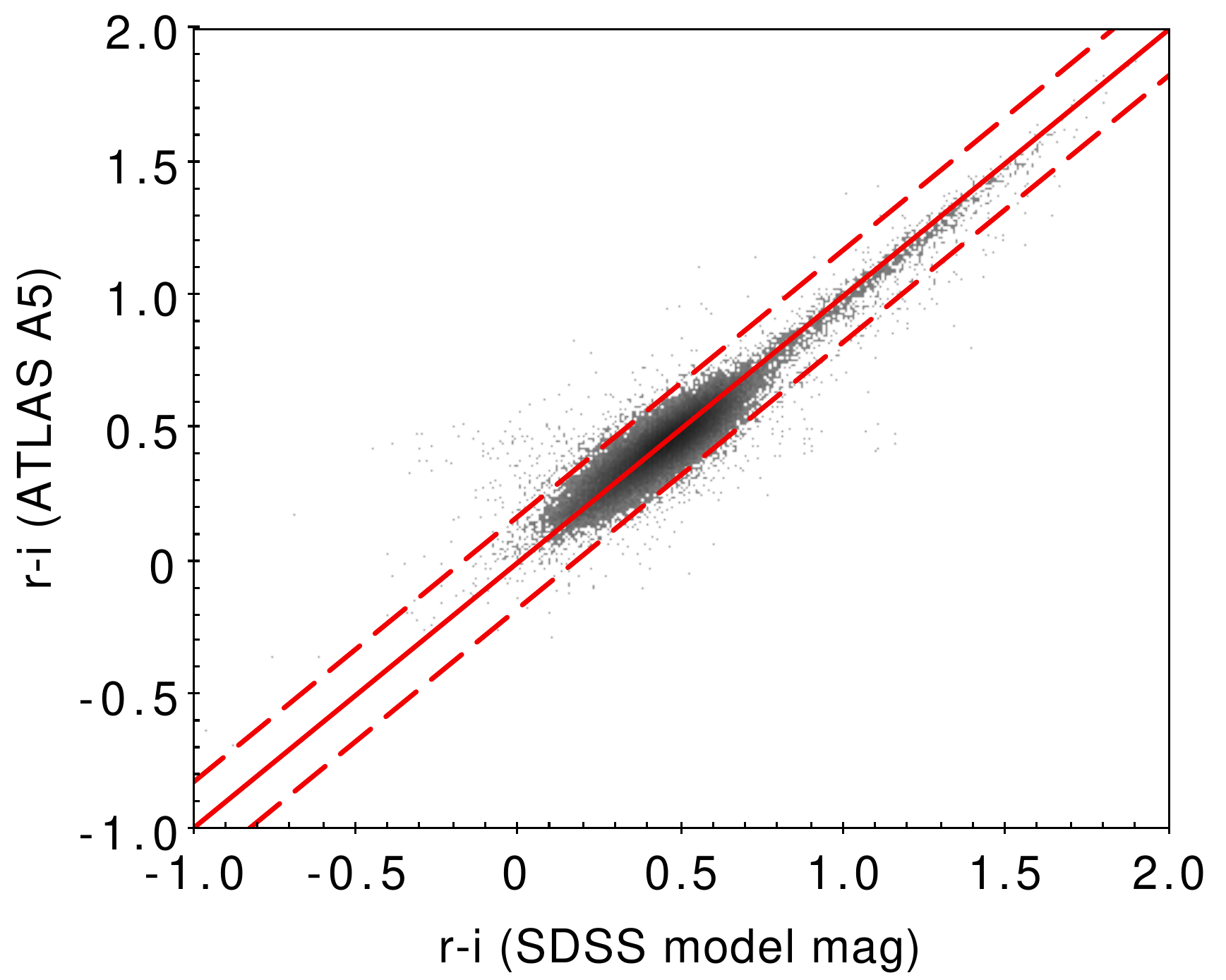} 
		\label{fig:r-i_ATLAS_SDSS}
	\end{subfigure}
		\begin{subfigure}{0.66\columnwidth}
		\includegraphics[width=\columnwidth]{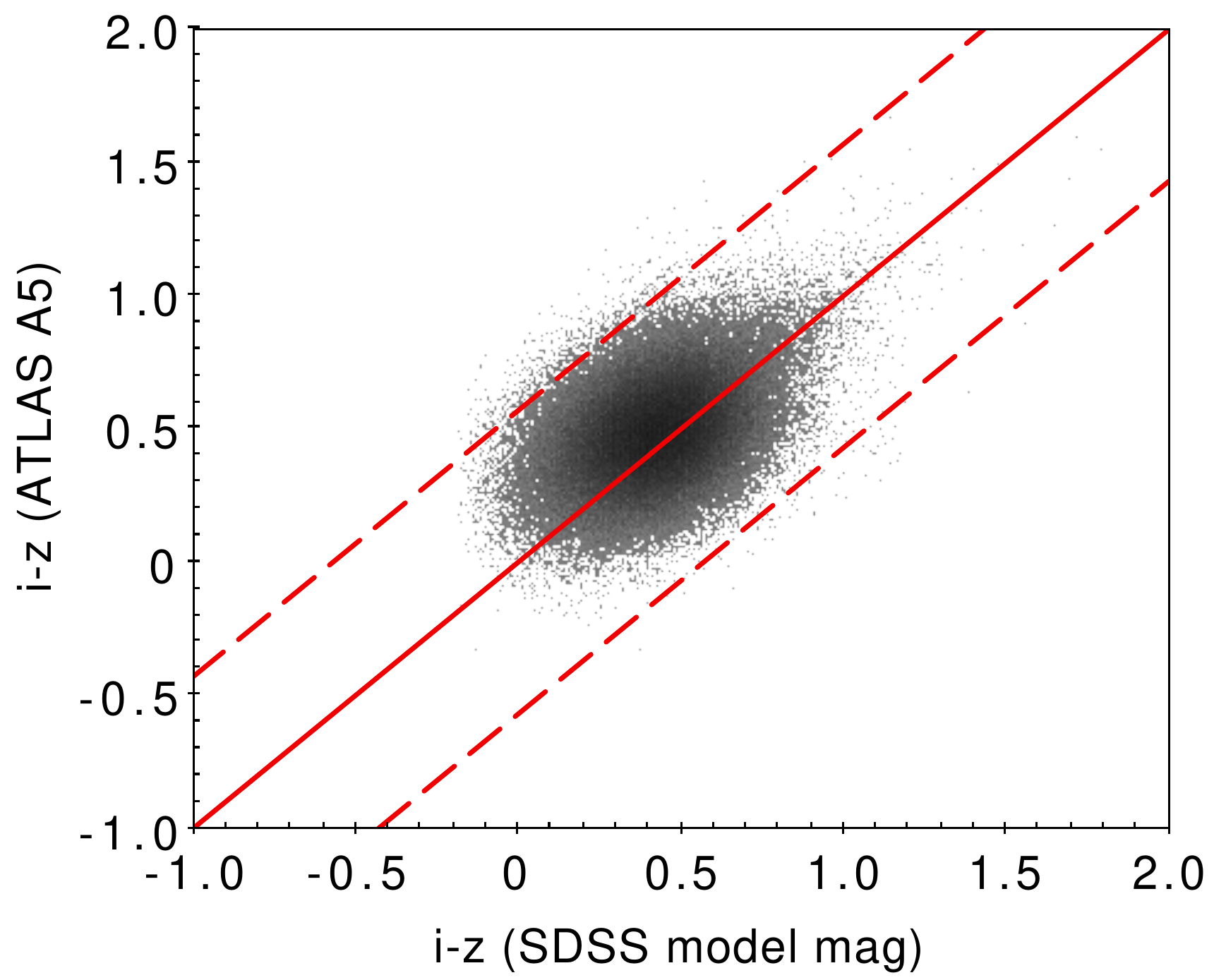} 
		\label{fig:i-z_ATLAS_SDSS}
	\end{subfigure}
	\caption[]{A comparison of the g-r, r-i and i-z colours based on VST ATLAS Aperture 5 and SDSS model magnitudes. Given the similarity between the colours, we use the SDSS cuts in our LRG sample selections, described in Section~\ref{sec:LRG_data}. Here the dashed lines indicate the $3\sigma$ outliers.}
	\label{fig:ATLAS_SDSS_colours}
\end{figure*}

Figure~\ref{fig:ATLAS_SDSS_colours} shows a comparison of VST ATLAS Aperture 5 and SDSS model magnitude colours. Upon removing $3\sigma$ outliers (as indicated by the dashed lines in the plots), we find a rms scatter of 0.03, 0.04 and 0.13, with ATLAS-SDSS offsets of 0.01, -0.01 and 0.05 for $g-r$, $r-i$ and $i-z$ colors respectively. Here we impose magnitude limits of $g<21.5$, $r<19.5$, $i<20.5$ and $z<20.0$, corresponding to the range of magnitudes used in our LRG selection. In each case we only impose the magnitude limits of the two bands used to obtain the colours and restrict the sample to galaxies based on the ATLAS morphological classification in those bands. Given the reasonably tight scatter and small systematic offsets, in this work we adopt a photometric selection criteria based on SDSS magnitudes, when defining our LRG samples.  

\section{LRG Contamination tests}
\label{sec:Appendix_B}

\begin{figure*}
    \begin{subfigure}[t]{\textwidth}
	\includegraphics[width=\textwidth]{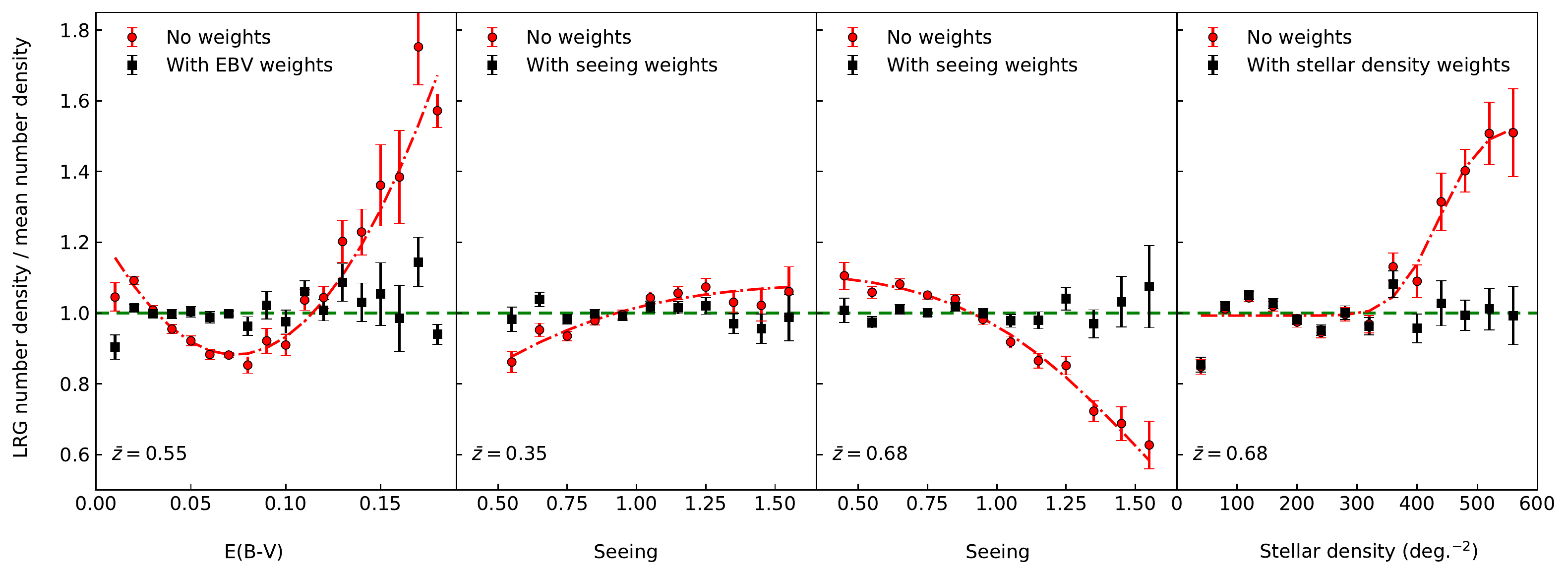}
	\label{fig:systematic_fits}
	\end{subfigure}
	\begin{subfigure}[t]{\textwidth}
	\includegraphics[width=\textwidth]{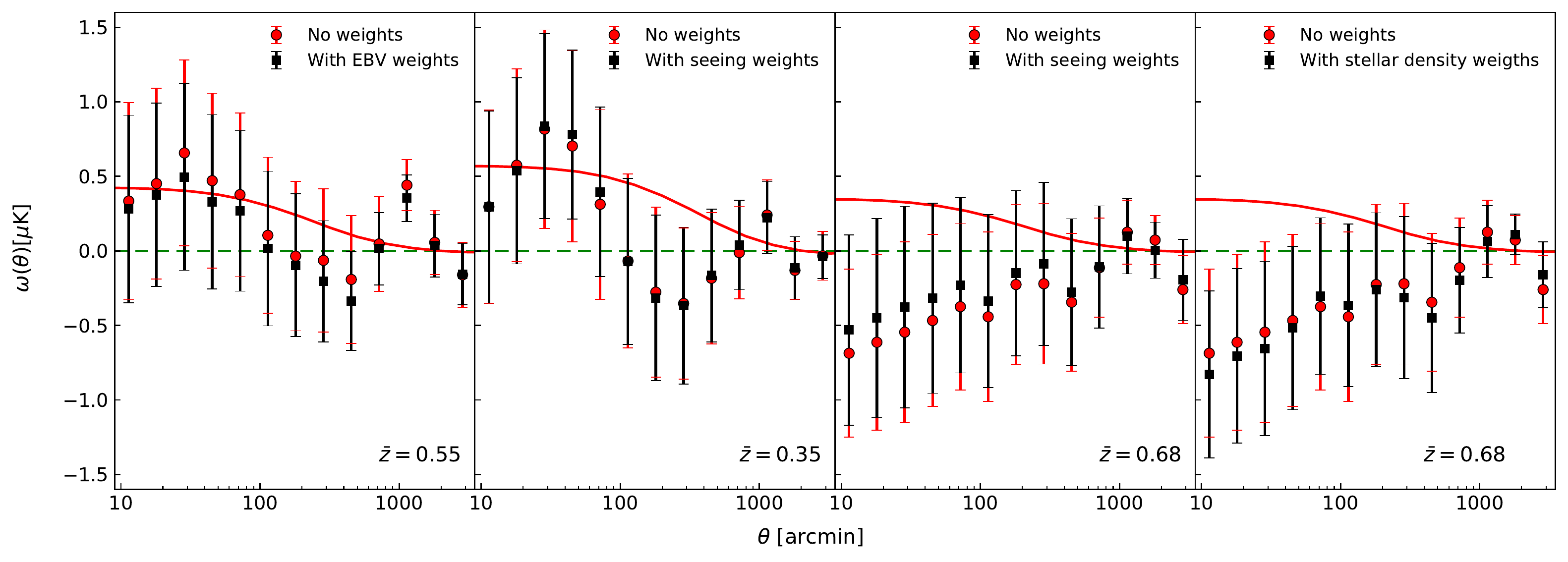}
	\label{fig:XCF_systematic_tests}
	\end{subfigure}
	\caption{Top panel: Normalised projected LRG number density as a function of Galactic extinction (in E(B-V)), seeing (in r-band for $\bar{z}=0.35$ and i-band for $\bar{z}=0.68$) and projected stellar density (limited to $19.8<i<20.5$ corresponding to the magnitude limits of the $\bar{z}=0.68$ LRG sample). Here the dot-dashed curves are the best-fit relationships used to define the weights correcting for the observed systematic trends. Bottom panel: The impact of including the E(B-V), seeing and stellar density weights from the top panel on our ISW measurements. In all cases, the inclusion of weights do not appear to have a significant impact on our ISW measurements.}
	\label{fig:systematic_tests} 
\end{figure*}

Adopting a similar approach to \cite{Ross2017}, we test for the impact of various sources of survey systematics including airmass, seeing, galactic dust extinction and stellar contamination on our $\bar{z}=0.35$, $0.55$ and $0.68$ LRG samples. In the top panel of Figure~\ref{fig:systematic_tests} we show the four instances where systematic trends due to galactic extinction, seeing and stellar contamination appear to be present in our LRG samples, finding no major systematic trends in the remaining cases. As shown in the bottom panel of Figure~\ref{fig:systematic_tests}, the inclusion of weights correcting for these observed systematics does not appear to have a significant impact on our ISW measurements.
%%%%%%%%%%%%%%%%%%%%%%%%%%%%%%%%%%%%%%%%%%%%%%%%%%

% Don't change these lines
\bsp	% typesetting comment
\label{lastpage}
\end{document}